\documentclass[preprintnumbers,nofootinbib,noshowpacs,eqsecnum,prd,superscriptaddress,letterpaper]{revtex4}

\usepackage[utf8]{inputenc}
\usepackage{graphicx}
\usepackage{amsmath,amssymb}
\usepackage{feynmp}
\usepackage{url}
\usepackage{ulem}
\usepackage{multirow}
\usepackage{hyperref,color}

\newcommand{\ttbar}{\ensuremath{t\bar{t}}}
\newcommand{\pt}{\ensuremath{p_\mathrm{T}}}
\newcommand{\pT}{\ensuremath{p_\mathrm{T}}}

\begin{document}

\title{Physics Goals and Experimental Challenges of the Proton-Proton
  High-Luminosity Operation of the LHC}

\author{P.\ Campana}
\affiliation{Laboratori Nazionali dell'INFN, Frascati, Italy}
\author{M.\ Klute}
\affiliation{Massachusetts Institute of Technology, Cambridge,
  USA}
\author{P.\ S.\ Wells}
\affiliation{CERN, Geneva, Switzerland}

\begin{abstract}
The completion of Run 1 of the CERN Large Hadron Collider has seen
the discovery of the Higgs boson and an unprecedented number of 
precise measurements of the Standard Model, while Run 2 operation has just started
to provide first data at higher energy. 
Upgrades of the LHC to high luminosity
(HL-LHC) and the experiments (ATLAS, CMS, ALICE and LHCb) will
exploit the full potential of the collider to discover and explore new physics beyond the Standard Model.
In this article, the experimental challenges and the physics opportunities in
proton-proton collisions at the HL-LHC are reviewed. 
\end{abstract}

\maketitle


\input{ar-hllhc.sty}

\section{Introduction}
\label{sec:intro}

The primary goal of the experiments at the CERN Large Hadron Collider (LHC) is to answer fundamental questions in particle physics. Several experimental collaborations use dedicated detectors to analyse the enormous number of particles produced 
at four interaction points around the accelerator.
ATLAS and CMS are general-purpose detectors to explore the vast physics landscape accessible at the LHC, while ALICE and LHCb are designed to focus on specific phenomena. The ALICE speciality is heavy-ion collisions, which are not considered here.

The highlight of Run-1 operation (2010--2013) was the observation of a Higgs boson~\cite{HIGG-2012-27, CMS-HIG-12-028} in 2012. However, the  experiments have only scratched the surface of the ultimate physics potential of the LHC. There are still many open fundamental questions in particle physics that can only be addressed by increasing both the energy  and the luminosity of the collider.

The first phase of LHC operation will conclude in 2023, when the original design luminosity should have been delivered. The High Luminosity LHC (HL-LHC) program will follow from 2026 
and promises ten times more data, significantly extending the reach of searches with the potential for major discoveries. 
The HL-LHC will probe the boundaries of the Standard Model (SM). The main pillars of the physics program are precision measurements of the SM including the Higgs boson, searches for new physics through the study of rare SM process, searches for new heavy states, and measurements of the properties of any newly discovered particles.

The accelerator upgrade relies on a number of key innovative and challenging technologies, such as cutting-edge superconducting final focus magnets in excess of 12~T and very compact and ultra-precise superconducting radio-frequency cavities. The experiments will also undergo major upgrades to fully exploit their potential. The substantial increase in luminosity will pose three major technical challenges. The expected average number of simultaneous proton-proton (pp) collisions (pile-up) will increase from 40 to up to 200, making each event much larger in size and much more complex to record and analyse. Faster detectors and readout electronics, as well as sophisticated  trigger systems to efficiently identify  physics signatures, will be required.  Finally, the detectors will need to tolerate a substantial radiation dose.

This review briefly summarizes the highlights of the LHC Run-1 physics
results, gives an overview of the detector and accelerator upgrades,
and describes the performance of the experiments for the HL-LHC pp
program with the main emphasis on physics opportunities.

\section{Highlights of Physics Results at the LHC}
\label{sec:run1}

During Run~1, ATLAS and CMS each recorded about 5~\ifb\ at $\sqrt{s}=7$~TeV, and 20~\ifb\ at 8~TeV. Focussing on flavour physics in the forward region, LHCb operated
at lower instantaneous luminosity and collected a total of 3~\ifb.
There were also periods dedicated to the heavy ion program with lead-lead or proton-lead collisions, and pp collisions at the corresponding nucleon-nucleon collision energies.

The standout result was the discovery 
of a new boson with a mass of about 125~GeV~\cite{HIGG-2012-27,CMS-HIG-12-028}. 
So far, this particle is
consistent with the SM Higgs boson with spin-0 and CP-even. The spin-2 hypothesis is ruled out
at more than 99.9\% confidence level (CL). 
The most precise mass measurements are from the $ZZ \rightarrow 4 \ell$ and $\gamma\gamma$ final states, with the combined ATLAS and CMS value of 
$m_H=125.09 \pm 0.21 (\mathrm{stat.}) \pm 0.11 (\mathrm{syst.})$~GeV~\cite{HIGG-2014-14}.
The results are comfortably in the range predicted by a combined fit
to electroweak measurements~\cite{Baak:2012kk}, but could also be a
plausible lightest CP-even, neutral Higgs boson in supersymmetric models.

The Higgs boson couplings are inferred from the basic measurements of
cross sections for different production modes times branching
fractions into different final states. 
These are expressed as signal strengths, $\mu$, normalised to the SM expectation.
Assuming SM ratios of cross sections and decays, the overall ATLAS and
CMS combined signal strength has a precision of about
10\%~\cite{ATLAS-CONF-2015-044}, while the strengths for the five main
decays to $\gamma \gamma$, $ZZ$, $WW$, $\tau \tau$ and $bb$ are
measured to 20--30\% precision. The bosonic final states are
established at greater than 5$\sigma$ significance by each experiment
alone. Combining the two experiments also brings the  $\tau \tau$ decay mode to this level.

The LHC Run~1 has provided a wealth of measurements~\cite{ATLASCMSall} of
electroweak and QCD process at 7 and 8~TeV, ranging 
from the total
cross section of about 80~$\mu$b to rare processes such as $\ttbar V$ production with a 
cross section of less than 1~pb. QCD jet production has been measured in
great detail, with the rate of jets as a function of $\pT$, and of dijets as a function of 
the jet-pair mass, spanning orders of magnitude. 
In the case of $W$ and $Z$ boson production, differential cross sections
such as the $\pT$ distribution, and the rates of up to 7 additional jets have been explored.
The precision of theoretical predictions have advanced in parallel, but there are
still improvements to be made, with discrepancies in the detailed modelling of
differential distributions even at NNLO, and uncertainties in predictions from 
scale variations and PDFs.
The abundant production of $\ttbar$ pairs and single top quarks has also allowed
precise measurements to be made.
The production of pairs of vector bosons is well established. Of particular interest is
electroweak production
of $W$ boson pairs by vector boson scattering, 
where there is first evidence of a signal~\cite{STDM-2013-06,Khachatryan:2014sta}.

LHCb has proven to be a true general purpose forward spectrometer, with measurements of the $W$, $Z$, and top cross sections. 
The measured forward top cross section will provide an additional
handle in reducing PDF uncertainties.
However, it is in the domain of flavour physics where the experiment made the most significant impact.
A highlight is the observation of the 
extremely rare $B^0_s \to \mu \mu$ decay at $6\sigma$ significance in a 
combined CMS and LHCb analysis \cite{bsmumu}.
In the CP violation sector, LHCb has measured the
CKM angle $\gamma$ with the best single experiment precision,
obtained from time integrated analyses of $B^+ \to Dh^+$ and $B^0 \to D K^{*0}$ decays \cite{gamma}.
The CP violating phase $\phi_s$ in  $B_s \to J/ \psi K^+ K^-$ and $B_s \to J/ \psi \pi^+ \pi^-$ processes has been determined \cite{phis}.
CP violation in the charm sector has been probed to the level of $10^{-4}$ \cite{cpv-c}, together with a precise measurement of the $D^0 \overline{D}^0$ mixing parameters \cite{mixing-c}.

With limited statistical significance, LHCb has observed anomalies with respect to SM predictions, that attracted theoretical interest.
These are the angular observables in $B^0 \to K^* \mu \mu$ decays~\cite{b2kstarmumu}, the ratio of $B^0 \to D^* \tau \nu$ to $B^0 \to D^* \mu \nu$ decays~\cite{tau-mu}, and the exclusive determination of 
$|V_{ub}|$ CKM matrix element using the $\Lambda_b \to p \mu \nu$ decay~\cite{vub}.
Further theoretical and experimental efforts are needed to confirm or disprove these effects.

LHCb measurements of  charm and onia central exclusive production in the forward region have been made, and spectroscopy studies have proved fruitful.
Two new penta-quark charmonium resonances, $P_c^+ (4450)$ and $P_c^+ (4380)$, were unambiguously identified \cite{p-quark}, the resonance nature of the tetra-quark $Z(4430)$ \cite{z}, 
and the quantum numbers of  the $X(3872)$ particle determined \cite{x}.

Many searches for signatures of physics beyond the SM have been performed.
In brief, no compelling hints
have emerged from Run~1 from ATLAS or CMS. Supersymmetric models motivate diverse
searches,
with limits on first and second generation squarks and
gluinos reaching nearly 2~TeV~\cite{ATLAS-SUSY-2014-06}.
The sensitivity to these strongly produced
SUSY particles increases rapidly with $\sqrt{s}$. Searches for 
rarer processes such as third generation squarks
require large integrated luminosities, as do searches for electroweak-inos 
and particles in compressed SUSY models~\cite{ATLAS-SUSY-2014-08}.

More exotic models of new physics have also been explored,
with one or two small excesses which will be scrutinised 
with more data in Run 2 (see for example~\cite{ATLASDiboson, Khachatryan:2014hpa}).
In this case, the
scale of new physics probed can be even higher, reaching above 20~TeV in
the extreme case of contact interactions. Again, 
searches for strongly produced objects benefit from increasing $\sqrt{s}$, 
such as quantum black holes in theories of extra dimensions, 
which have very high pp production cross sections.
New particles such as vector-like-quarks
which would provide an alternative way of cancelling top-quark contributions
to the Higgs mass have been explored up to nearly 1~TeV. 
Exotic signatures also include generic dark matter candidates, which
can be detected if they are produced together with  
initial state radiation of a jet or additional particles like vector
bosons or photons. These searches are complementary to
direct particle-astrophysics searches for the unexplained dark matter
in the universe.

The year 2015 saw the beginning of Run~2 at 13~TeV, close to the
full design energy of 14~TeV. By the end of the year,
LHCb collected about 400~\ipb, with the first results on $b$ and
$c$-quark production cross sections~\cite{run2-b, run2-c}.
A plethora of preliminary results were available from 
ATLAS and CMS with 2 to 3~\ifb\ of pp data.
In addition to the total inelastic cross section,
many production rates were
measured to be in broad agreement with the SM, 
including inclusive jets, 
$W$ and $Z$ bosons (with jets), dibosons, $\ttbar$ and ($t$-channel)
single-top.
Searches, in particular for strongly produced SUSY or exotic processes, were already
sensitive beyond the Run~1 limits. 
A small but intriguing excess of diphoton
events with a mass around 750~GeV provoked much 
speculation~\cite{ATLAS750,CMS750}, and more
data are eagerly awaited.

\section{Accelerator and Detector Upgrades}
\label{sec:detectors}

\subsection{LHC Upgrades}

The LHC alternates ``Runs'' of a few years' operation in a relatively stable configuration, with long shutdowns (LS) in which major work may be carried out on the accelerator and experiments. After consolidation work in LS1, the machine is now able to reach close to the 14~TeV design energy. During LS2, in 2019--2020, the Phase-1 injector and cryogenic upgrades will allow the luminosity to reach about twice the nominal value of $10^{34}$~cm$^{-2}$s$^{-1}$, to deliver
the original target of 300~\ifb\ by the end of Run 3 in 2023.
At this point the final focussing magnets in the interaction regions need to be
replaced, which gives the opportunity to replace them with stronger focussing magnets in order 
to increase the instantaneous luminosity.

The main objectives of the HL-LHC project~\cite{Barletta2014352} are to operate with a peak levelled luminosity of 5 (baseline) to 7.5 (ultimate) $\times 10^{34}$ cm$^{-2}$s$^{-1}$ and to deliver integrated luminosities of 250~\ifb\ per year enabling the goal of  3000~\ifb\ within a twelve year time frame. 
The corresponding Phase-2 upgrade of the
HL-LHC interaction region will require quadrupole magnets with an increased aperture of 150~mm and peak field in excess of 12~T, beyond the capabilities of the NbTi conductor. Additional upgrades will be needed to the injector systems, radio frequency cavities, 
collimators, and
high temperature superconducting links.

\subsection{ATLAS Detector}

The plans for the ATLAS Phase-2 
upgrades \cite{CERN-LHCC-2012-022, CERN-LHCC-2015-020} represent a coherent
progression from the consolidation program completed in the LS1 shutdown
and the Phase-1 detector configuration to be deployed during LS2,
in order to maintain excellent performance with increasing instantaneous
luminosity. The Phase-1 upgrades are designed with Phase-2 compatibility in mind.
The major change
will be 
the complete replacement of the inner tracking detector.
Many additional upgrades are required to improve the trigger capabilities, and 
sustain the higher event rates 
needed to record all the events of interest despite the ten times higher
instantaneous luminosity.


The present ATLAS design has a hardware Level-1 (L1) trigger and higher
level triggers (HLT) implemented in software.
A two-level hardware trigger is foreseen with the new Level-0 (L0)
trigger using calorimeter and muon information.
The L0 and L1 triggers are designed to operate at rates up to
 1~MHz and 400~kHz, respectively.
The L0 calorimeter triggers will be largely based on the Phase-1
L1 calorimeter trigger system~\cite{ATLASPH1TDAQ} and LAr trigger
electronics~\cite{ATLASPh1LAr}, which  improve the granularity of the
calorimeter information by a factor of ten. Upgrades to the 
muon triggers are planned to provide the L0 trigger.
A fast track trigger (FTK)~\cite{ATLASFTK} is being
installed in Run 2, using
hardware track reconstruction based on pattern recognition in associative
memory chips to reconstruct tracks for use in the HLT.
Similar solutions will be used in the Phase-2 upgrade both at L1 and 
in the HLT. The HLT event filter processors
have a planned output rate of 10~kHz.
The computing and software of the experiment
will also evolve as needed, incorporating new computing architectures.


In Run 1, the inner detector included silicon pixel and microstrip layers, 
and a straw-tube transition radiation tracker. 
An additional innermost pixel layer~\cite{ATLASIBL}, was installed
during LS1.
For reasons of high occupancy and radiation damage, the tracker will be completely replaced for HL-LHC operation.
The baseline design inner tracker (ITk)~\cite{CERN-LHCC-2012-022, CERN-LHCC-2015-020}
is an all-silicon detector, with pixel layers at the inner radii, surrounded by microstrip sensors. 
The cylindrical central region has
4 pixel layers and 5 microstrip layers, with smaller pixel sizes and where necessary shorter
strip lengths than the present tracker to maintain suffciently low occupancy.
The pixel endcaps extend in the very-forward region 
to cover  $|\eta|=4.0$, and there are 7 strip disks.
Its performance will be enhanced by a 
lower mass construction, reducing the effect of multiple scattering, photon conversions and hadronic interactions. 
The detector will be immersed in the 2~T magnetic field provided by the present ATLAS solenoid,
and will provide improved impact parameter and momentum resolution.



The ATLAS calorimetry uses liquid argon detector technology for the
electromagnetic barrel and endcap, the hadronic endcap and the
forward calorimeters, and scintillating tiles for
the hadronic barrel calorimeter.  
The Phase-2 upgrade foresees new forward calorimeters with higher transverse granularity for improved
handling of the large fluctuations in the energy deposition due to the high pile-up.
It is also planned to introduce a finely segmented precision timing
detector in front of the endcap calorimeter, covering approximately the range $2.4 < |\eta| < 4.3$,
to improve the identification and rejection of pile-up energy clusters.
The readout electronics of the remaining calorimeters will
be upgraded for radiation tolerance and to operate at the L0/L1 trigger rates.
Information from the last layer of the tile
calorimeter will be included in the muon trigger to
assist in rejecting fake muons.


The muon spectrometer is immersed in large air-core superconducting 
toroidal magnet systems, two endcaps and one barrel, providing a 
field of about 0.5 T.
Muon trajectories are measured by three stations
of precision monitored drift tube (MDT) chambers over most of the acceptance.
Resistive plate chambers (RPC) in the barrel, and thin gap chambers (TGC)
in the endcap big wheels provide the hardware trigger.
New Small Wheels~\cite{ATLASPh1NSW} will replace the first layer of the endcaps
in the Phase-1 upgrade, using
micromegas chambers for tracking, and small strip TGC (sTGC)
for triggering.
The Phase-2 plans include upgrades to the trigger and readout electronics
to support higher occupancies and trigger rates.
Additional RPC and sMDT chambers will be installed in the inner barrel
to cover acceptance holes, and compensate for RPC 
longevity issues.
Trigger chambers in the forward region $2.0<|\eta|<2.4$
will be replaced with sTGC to improve pile-up rejection.
Finally, the muon acceptance will be increased with a very forward tagger
covering $2.6<|\eta|<4$ to match the inner tracker extension.

\subsection{CMS Detector}

The proposed CMS detector upgrades~\cite{CERN-LHCC-2015-010} are mandatory to mitigate effects on the physics performance from the HL-LHC beam conditions and damage from radiation. The most significant upgrades are the replacement of the tracking detector and the endcap calorimeters. 

The individual components of the CMS Phase-2 upgrade proposal are discussed briefly in the following.  

The Phase-2 outer tracker and pixel systems are designed with roughly a factor four larger granularity to maintain adequate track reconstruction performance at the much higher pile-up levels of the HL-LHC. This will be achieved by shortening the lengths of silicon sensor strips relative to those in the current detector, without changing the pitch very significantly. A number of design improvements will lead to a much lighter outer tracker providing significantly improved $\pt$ resolution and a lower rate of $\gamma$-conversions, compared to the present detector. In addition, the module design will be capable of providing track-stub information to the L1 trigger at 40 MHz for tracks with $\pt \ge 2$~GeV. This will ensure powerful background rejection at the earliest stage of the event selection. The pixel system will implement smaller pixels and thinner sensors for improved impact parameter resolution and better two-track separation. With ten pixels disks in each of the forward regions the system coverage will be extended to close to $|\eta| = 4$.

The endcap calorimeter will consist of a High Granularity Calorimeter (HGC) including both an electromagnetic section and a hadronic section, followed by a rebuilt conventional hadronic endcap (HE) with reduced depth. The design of the HGC uses silicon sensors as the active material with pads of variable sizes of less than a $\mathrm{cm}^2$. In the longitudinal direction, 30 planes form an electromagnetic section with $25X_0$ and one nuclear interaction length ($\lambda$). The next 12 planes with a depth of 3.5$\lambda$ cover the hadronic shower maximum measurement, and are followed by an HE of similar design to the current detector to provide an additional depth of 10$\lambda$. 

The upgrade of the endcap muon system is designed to increase the redundancy in the region $1.5 \le |\eta| \le 2.4$. The two first stations are in a region where the magnetic field is still reasonably high. Gas Electron Multiplier (GEM) chambers will be used for good position resolution in order to improve momentum resolution for the standalone muon trigger and matching with tracks in the global muon trigger. Low-resistivity  Resistive Plate Chambers (RPC)  will be used for the two last stations will use with lower granularity but good timing resolution to mitigate background effects. In addition, the implementation of a GEM station in the space that becomes free behind the new endcap calorimeters is being proposed in order to increase the coverage for muon detection to $|\eta| \approx 3$. 

The reconstruction and matching of charged particles to muon and calorimeter information by the L1 trigger system requires an increase of the latency from 3.4~$\mu$s to 12.5~$\mu$s. This change will require upgrades of the front-end electronics in some of the existing sub-detectors that will be kept for Phase 2. 
Based on the expected performance of the trigger with track information, a L1 trigger acceptance rate of 500 kHz for beam conditions yielding 140 pile-up events is proposed. This will allow to maintain thresholds comparable to those that will be used in a typical Phase-1 trigger menu. To retain comparable performance in beam conditions that result in 200 pile-up events, the L1 rate must increase to 750 kHz, and so all detectors will have readout capabilities compatible with this alternative. Any further increase of the L1 readout rate would require an increase of the pixel detector readout bandwidth. The upgrade of the front-end electronics for calorimeter and muon sub-detectors are required to meet these trigger requirements.

The Data Acquisition (DAQ) system will be upgraded to implement the increase of bandwidth and computing power that will be required to accommodate the larger event size and Level-1 trigger rate, and the greater complexity of the reconstruction at high pile-up. Compared to Phase 1, the bandwidth and the computing power requirements would respectively increase by factors of about 15 and 30 for operation at pile-up of 200. This is well within the projected network and computing technology capabilities expected at the time of Phase 2. 

\subsection{LHCb Detector}

LHCb has planned  an upgrade for 2021~\cite{upgrade}, with two crucial improvements: 
the capability to exploit a 5 times higher instantaneous luminosity ($\sim 2 \times 10^{33}$~cm$^{-2}$~s$^{-1}$), and a readout to acquire events at the LHC bunch crossing rate ($\sim$ 40~MHz). The upgrades enable the improvement of the precision of most of its key  measurements substantially during LHC Run 3 by the end of 2023, planning to integrate $\sim 50$~\ifb\ by 2030. 
Minimal modifications are needed to the LHCb interaction region to be compliant with the HL-LHC operation.

The collaboration is preparing to upgrade the detectors and the readout architecture to enable the transition of the trigger rate from the current 1~MHz to 40~MHz.

Events at a pile-up of $\sim 6$ are very crowded and contain hundreds of particles  from several primary vertices in the envelope of the effective LHCb pp collision zone which is a few cm long. A high granularity vertex detector (VELO) is required to handle these events. Therefore the current VELO will be replaced~\cite{velo}, with 50x50 $\mu \mathrm{m}^2$ silicon pixels to provide a full 3D reconstruction of tracks in space, and to limit the fake tracks. The upgraded detector also achieves the necessary radiation resistance, and it will have the first sensitive element at a radial distance of 4 mm from the beam axis.

The driving choices for the upgrade of the tracking system~\cite{trk} are a uniform and high granular readout over a large area ($\sim$ 100 m$^2$ per layer), a fast signal formation, and radiation hardness. A new approach, applied on a large scale for the first time, will be used: 2.5 m long modules made of 0.25 mm diameter scintillating fibres along the $y$ vertical direction, with an  $x$ coordinate resolution of $\sim 70~\mu$m. Fibres show excellent timing resolution and signals are contained in 25~ns. Fibre readout is performed with Silicon Photo Multipliers (SiPMs), sensitive to a single photon, that have a very high quantum efficiency. The temperature and voltage of the SiPMs must be controlled for a stable operation, while low temperature conditions must be met (below $-40\,^{\circ}{\rm C}$) to mitigate radiation damage.

A large data rate will be produced by the front-end electronics of the upgraded LHCb, up to 3 TB/s. To reduce the cost of the DAQ and computing infrastructures, a mix of state-of-the art technology solutions are under development \cite{daqtrg}: new generation of FPGA with enhanced computing and memory resources for the DAQ system, high rate network  to provide  bandwidth for data collection, and an Event Filter Farm (EFF) with massive  processing capacity provided also by graphic processing units (GPU). The  goal is to convey all the information from sub-detectors related to the same event into a processing unit, where the High Level Trigger (HLT) software will analyse and select the events. Due to the large  cross sections and the high trigger efficiencies, the output data rate will be significant. Data will be written to disk at a rate of 2 to 5~GB/s, with an average event size of $\sim$ 100~kB, which is challenging in term of computing resources for end user analysis. The collaboration is studying and testing new offline models in Run 2, also considering future developments in information technology.

Other systems \cite{pid} are refurbishing or modifying their layout to comply with the 40~MHz readout. The calorimeter and the muon systems will keep the current detectors but will upgrade part of the electronics, the silicon tracking system located between the vertex detector and the magnet will have new upgraded silicon strip sensors with a better granularity, while the two RICH detectors will replace the photon sensors with multi-anode PMs with a new front-end electronics.

\section{Performance with High Luminosity}
\label{sec:perf}

The experiments aim to measure all final state particles
such as electrons, photons, muons, tau leptons, 
and jets of hadrons. It is key to measure the impact
parameters with respect to the primary vertex and to reconstruct displaced
vertices to identify charm and bottom hadrons. The reconstruction of missing
transverse energy allows to the presence of 
neutrinos and possibly new weakly interacting particles to be inferred.
With increasing luminosity,
pile-up mitigation is achieved by combining 
detector upgrades and algorithm developments. 

\subsection{General Purpose Detectors}

In addition to maintaining 
excellent detector performance for all kinds of final state particles at HL-LHC instantaneous luminosities, 
the general purpose detectors, ATLAS and CMS, aim to continue to trigger on sufficiently low $\pT$ signatures as to be able to study electroweak processes or Higgs boson decays. 
Precise charged track and vertex reconstruction are a fundamental starting point. The impact of the pile-up from on average 140 (design) or 200 (ultimate) lower energy events can be controlled if final state particles can be linked to the primary vertex which matches the hard scatter event of interest. Information from the inner tracking detectors is also vital to identify charged leptons, converted photons, and to tag jets from heavy-flavour quarks, in particular b-jets.
Track reconstruction must also be efficient in dense environments such as high energy jets. Both detectors use tracking information to improve the measurement of jet energies and of missing transverse energy, and the addition of high granularity precise timing information gives an additional handle for pile-up mitigation. 

The collaborations prepared comprehensive studies of the detector performance in fully simulated samples with a variety of signals and with high 
pile-up~\cite{CERN-LHCC-2012-022,CERN-LHCC-2015-020,CERN-LHCC-2015-010,CERN-LHCC-2015-019}.
The relatively robust algorithms adapted from Run 1 and Run 2 data taking demonstrate that the Phase-2 upgrades deliver the required performance. It is inevitable that more sophisticated algorithms will be developed in future. 

\subsubsection{Tracking and vertex reconstruction}

The experiments have demonstrated that the track finding efficiency for pions remains across the extended pseudorapidity range of central trackers, using samples of $\ttbar$ events with up to 200 events average pile-up, where pions constitute the majority charged particles produced, as shown for example in Figure~\ref{fig:CMStrackbtag} (left).
The efficiency is about 90\%, while keeping the rate of fake tracks under control, i.e.\ misreconstructed tracks or tracks from random combinations of hits in the detectors. Since muons do not suffer from hadronic interactions in the detectors, the efficiency to reconstruct a muon track is uniformly higher and above 98\%.

\begin{figure}
\includegraphics[width=0.54\linewidth]{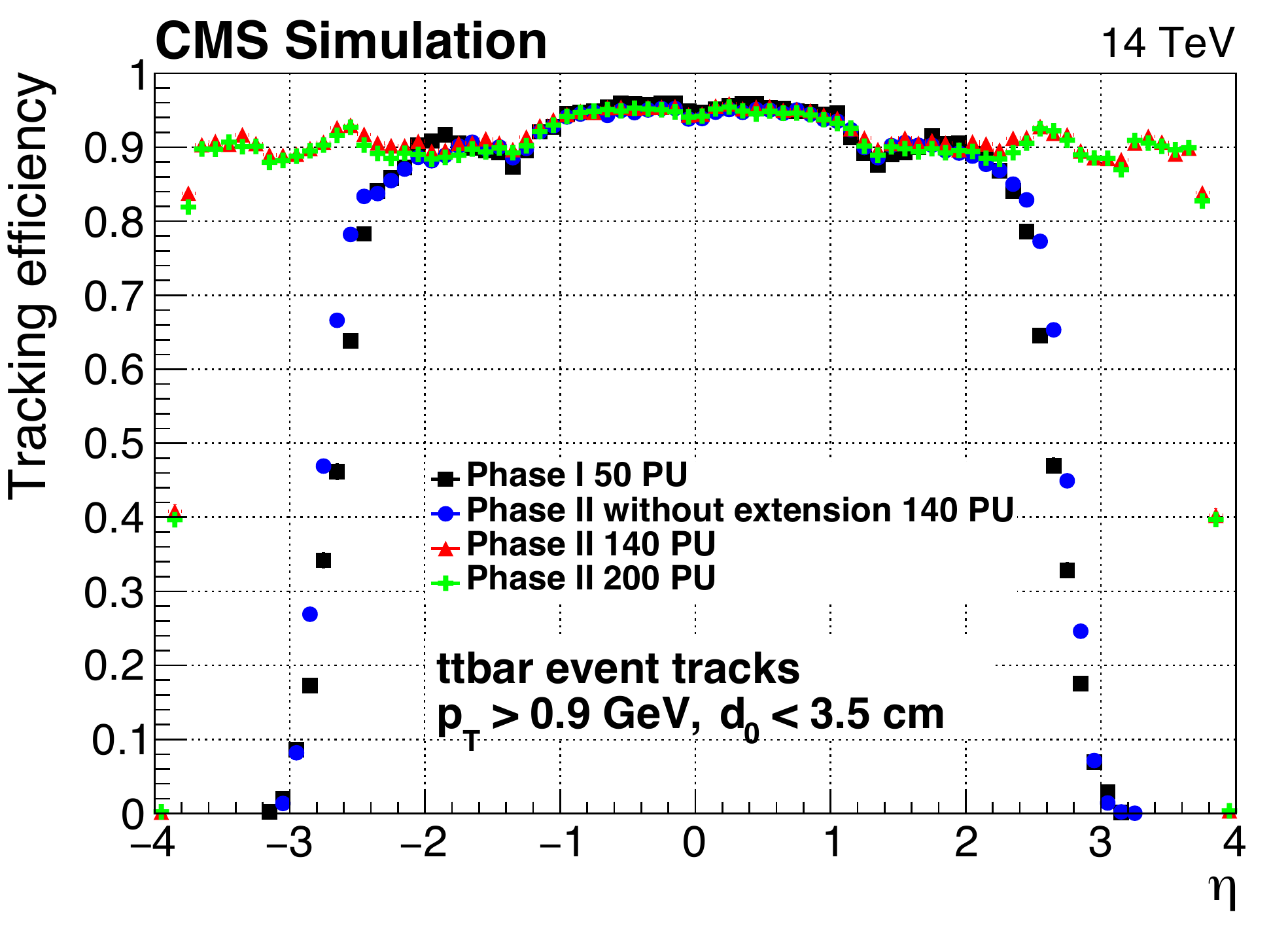}
\includegraphics[width=0.44\linewidth]{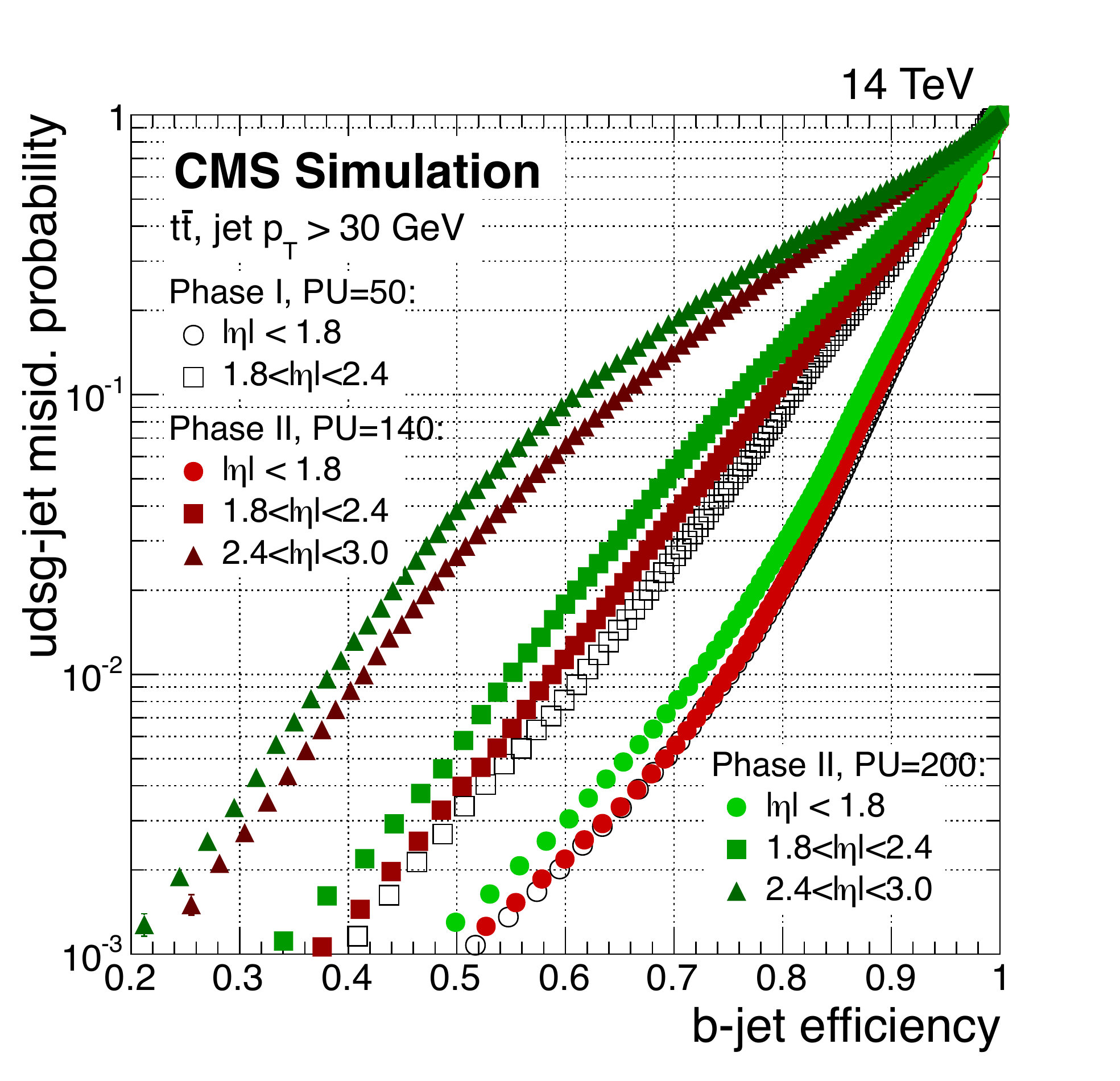}
\caption{(Left) Tracking efficiency in $\ttbar$ events for different CMS tracking detector configurations. (Right) The rate of identifying light-flavour jets as a b-jet, as a function of the b-jet tagging efficiency, as the tagging algorithm working point is varied. The sets of curves are for different ranges of $\eta$, and the colours indicate which detector layout and pile-up rate is assumed.}
\label{fig:CMStrackbtag}
\end{figure}

The primary vertex reconstruction efficiency is also very high for $\ttbar$ events, in the region of 98\%, and even a simple algorithm such as selecting the reconstructed vertex with the highest sum of transverse momentum squared of the associated tracks is able to ensure the selection the correct primary vertex. 
For lower multiplicity final states, such as inclusive $Z \rightarrow \mu^+ \mu^-$ events, or Higgs boson production by vector-boson fusion, where the Higgs boson decays to a photon pair, the vertex reconstruction efficiency drops by only a few percent.  The chance of identifying the correct vertex can be 10--20\% lower. The excellent impact parameter resolution, driven by the small pixel size of the innermost tracking layers, gives a 
longitudinal position resolution of 10--15~$\mu$m, and nearby vertices are typically not merged in if they are more than about a few hundred microns apart,
which matches the expected longitudinal density.

\subsubsection{Tagging of b-jets}

Jets with heavy flavour decays can be identified from the relatively long lifetime of charm and in particular bottom hadrons. The same excellent tracker performance required to reconstruct primary vertices also allows for efficient b-tagging with good rejection of other jets, including pile-up jets. ATLAS and CMS demonstrated that b-tagging algorithms with the new tracking systems perform well with an average pile-up of 140. There is a few percent decrease in b-tagging efficiency for the same rate of mistagging light-flavour jets when increasing from 140 to 200 events pile-up, see Figure~\ref{fig:CMStrackbtag} (right).
Further improvements in performance are expected as the algorithms are tuned, more 
sophisticated algorithms are applied, and the detector layouts are optimised.

\subsubsection{Lepton and photon identification}

Studies are in progress to further improve the electron, photon, muon and tau identification with the new layouts. In all cases, the use of isolation requirements to identify prompt leptons or photons is sensitive to the presence of pile-up, and benefits from algorithms to distinguish between activity from the hard scatter event and from pile-up interactions. There is a more noticible degradation of identification efficiency and energy or momentum resolution for electrons, photons and tau-leptons. However the reduced material in the upgraded detectors reduces the rate of photon conversions in the tracker volume. 

As discussed above, the efficiency to find a muon track in the central part of the detectors is very high. Matching to a track segment in the muon spectrometer is only weakly affected by pile-up, so for muons the main issues are related to maintaining a high trigger efficiency with low fake rate, and the impact of isolation variables.

\subsubsection{Jet reconstruction and missing transverse energy}

Particles from pile-up events potentially make a significant contribution to the measured energy of low $\pT$ jets from the hard scatter event of interest. For a pile-up of 140, the average additional energy from pile-up in a jet of radius $R=0.4$ in $\eta$--$\phi$ space is around 40~GeV, increasing to 60~GeV for a pile-up of 200. In addition to random combinations of particles from different pile-up events creating ``fake'' jets, there can also be genuine jets from the additional interactions in the event. 

The reconstructed jet energy depends on detector specific algorithms. The true jet energy is derived by applying a jet energy scale correction. 
Increasing the effective noise threshold for energy in a calorimeter cell suppresses low energy contributions from pile-up. Other techniques have been developed to estimate and remove or subtract the contribution from pile-up. For example, in running to date, ATLAS developed algorithms to estimate the pile-up in the event, and subtract an amount proportional to the jet area. The algorithm uses information on the sum of transverse momentum of tracks associated with the jet and originating from the primary vertex to decide if a jet is from the hard scatter or a pile-up jet. By default, CMS reconstruct jets from energy flow objects, which already combine tracking and calorimeter energy information. Several algorithms to mitigate against pile-up have been investigated of which the most effective is PUPPI~\cite{Bertolini:2014bba}. 

Figure~\ref{fig:ATLjetmet} (left) shows the rate per event of pile-up jets with $\pT > 30$~GeV as a function of $\eta$ for the ATLAS experiment. The majority of these jets can be removed by applying a track-based pile-up jet rejection algorithm, emphasising the importance of increasing the tracker coverage in the forward region.

\begin{figure}
\includegraphics[width=0.49\linewidth]{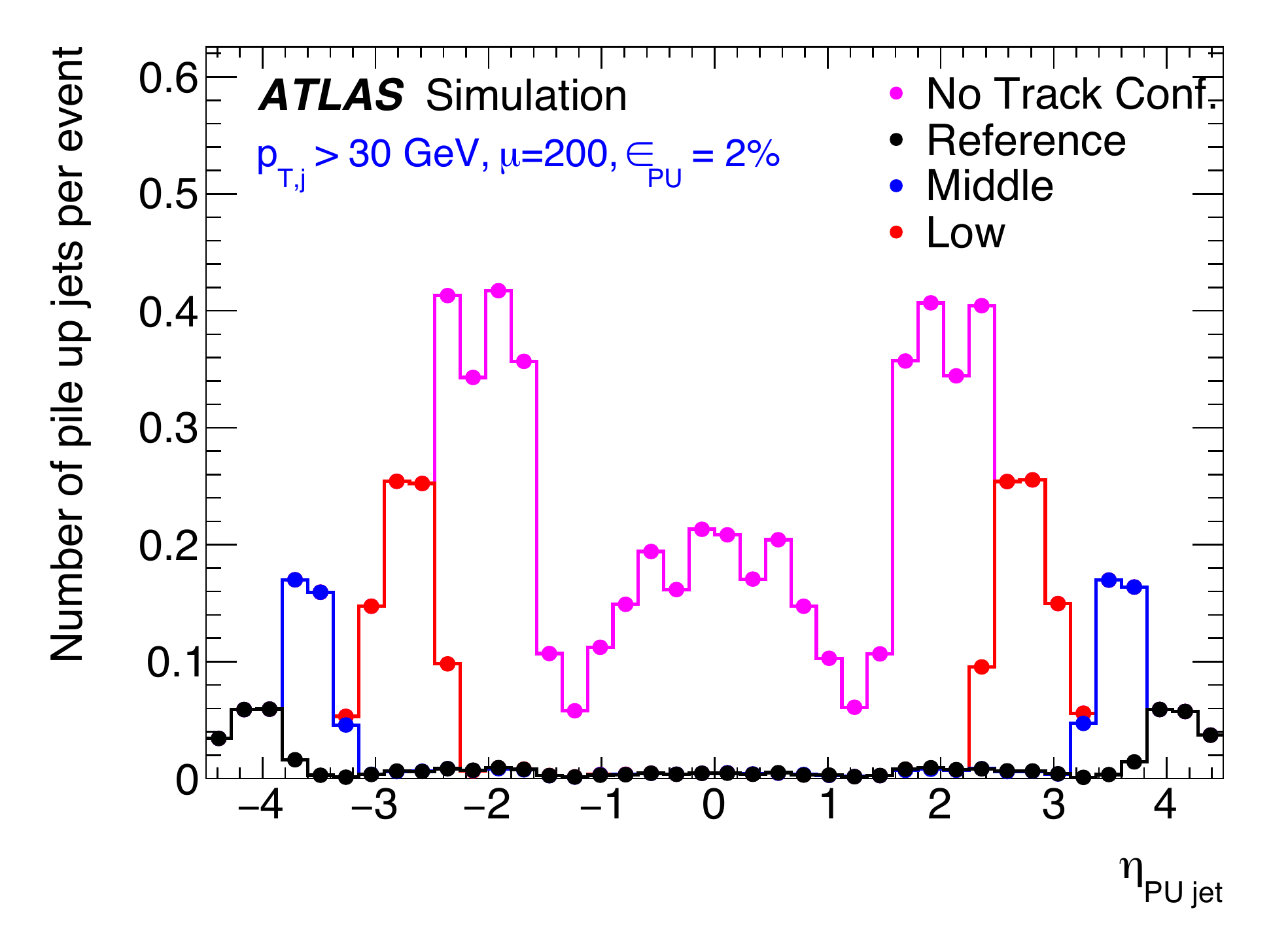}
\includegraphics[width=0.49\linewidth]{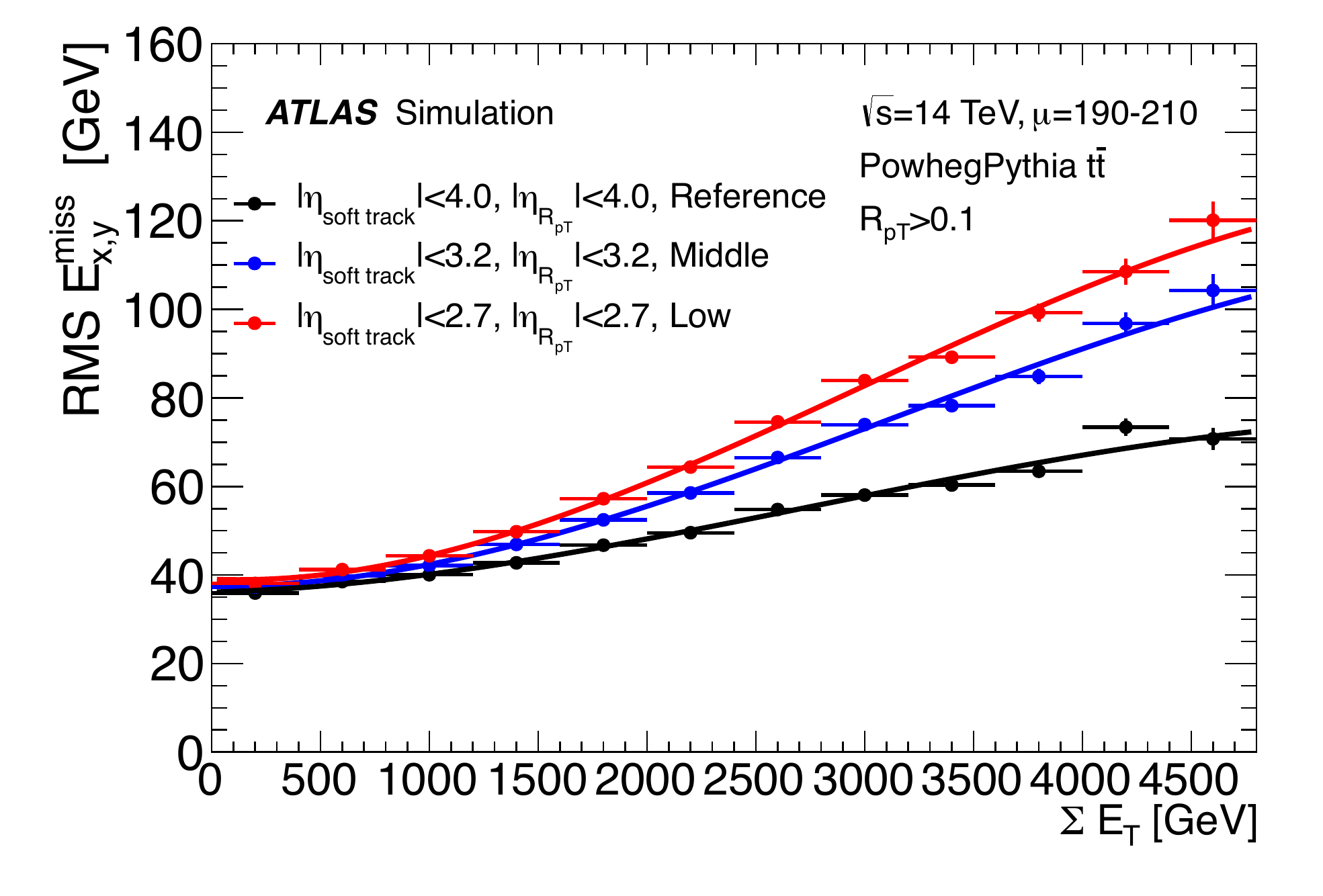}
\caption{(Left) The rate per event of jets with $\pT >30$~GeV from additional interactions wtih an average pile-up of 200. (Right) The \MET\ resolution for pile-up of 200 in three different pseudo-rapidity ranges.}
\label{fig:ATLjetmet}
\end{figure}

The missing transverse momentum vector in an event is the negative sum of the transverse momenta of all the measured particles, and its magnitude is the missing transverse energy, \MET. A significant \MET\ value indicates that one or more high energy non-interacting particles have been produced. These could be neutrinos, but could also be a new particle such as a dark matter candidate. Contributions from pile-up degrade the resolution of the \MET\ measurement. In Figure~\ref{fig:ATLjetmet} (right), the transverse energy resolution as a function of the scalar sum of transverse energy in the event is shown for a high pile-up scenario, and three different pseudo-rapdidity coverage ranges. The added benefit of forward tracking to reject pile-up jets is clear.

\subsection{LHCb}
The reconstruction of vertices, the tracking, the particle identification, and the triggering at a quality
comparable with  Run 1 and 2, represent the main challenges for LHCb at high luminosity.

\subsubsection{Vertex reconstruction}
The VELO reconstructs displaced vertices, and plays a significant role in the trigger and tracking. At trigger level it identifies tracks with high impact parameter (IP) and represents the first handle to reduce the background from minimum bias events. An excellent secondary vertex resolution is vital to resolve the fast $B^0_s$ oscillations.

Key performance parameters include the occupancy, the hit resolution, the track reconstruction efficiency and the decay time resolution. The occupancy in the pixel detectors reaches a maximum of 0.12~\% at the innermost sensor. The single hit resolution is $\sim$12$\mu$m, and the pattern recognition has a reconstruction efficiency of $\sim$99 \% for tracks with $p>5$~GeV. 
The 3D tracking capability of a pixelated layout is superior to the current system.

The IP resolution has a strong dependence on the $\pT$ of the track, while the primary vertex (PV) position resolution depends on the number of tracks belonging to the vertex. The upgrade performance is very similar to the present one. As a figure of merit, at $\pT \sim 1$~GeV a 3-dimensional IP resolution of 50 $\mu$~m is obtained,  see Figure~\ref{FIG1} (left), while standard deviations of 10 and 50 $\mu$m are observed in the $x$ and $z$ coordinate of the PV resolution, respectively. A decay time resolution  of $\sim 40$~fs is obtained for the $B^0_s \to J/\psi K K$ channel, equivalent to the current performance. The VELO is radiation hard up to 50~\ifb, without visible effects on performance.

\subsubsection{Tracking performance}
The most valuable tracks for physics analysis
are those (defined as {\it{long}}) which are reconstructed in the VELO and in detectors before (UT stations) and after (T stations) the magnet. 
{\it{Long}} tracks have excellent spatial resolution and a precise momentum measurement due to the combined information of the track slopes before and after the magnet.
Moreover, there are tracks  ({\it{downstream}}) measured only in the UT and in T stations,
reconstructing long-lived particles which decay outside the VELO
($K_{s}^{0}$ mesons and $\Lambda$ baryons) .

In the upgrade, the track reconstruction efficiency at high luminosity is  evaluated to be 99.6~\% for {\it{long}} tracks with $p > 5$~GeV, a value nearly identical to the current one. Thanks to the 3D reconstruction in the VELO,  the ghost rate is 2.5 \%, reduced by a factor 2 with respect to the current one.
For {\it{downstream}} tracks the situation is currently less satisfactory: the reconstruction efficiency in the upgraded detector is  lower by about 10\% (in absolute value) with respect to the current one and a higher ghost rate is observed.
The lack of {\it{y}}-segmentation in the fibre tracker is responsible for a part of the inefficiency, while the new algorithm is not yet optimal.
A re-design of the  inner zone is currently being investigated, introducing a  {\it{y}}-segmentation in a small area around the beam pipe, where the inefficiency is localized.

The  momentum resolution of the upgraded LHCb tracking system is very similar ($dp/p \sim 0.004$) to the current one, see Figure~\ref{FIG1} (right), and constant over a large momentum interval (up to $p \sim 50$~GeV).

\begin{figure}
\includegraphics[width=0.42\linewidth]{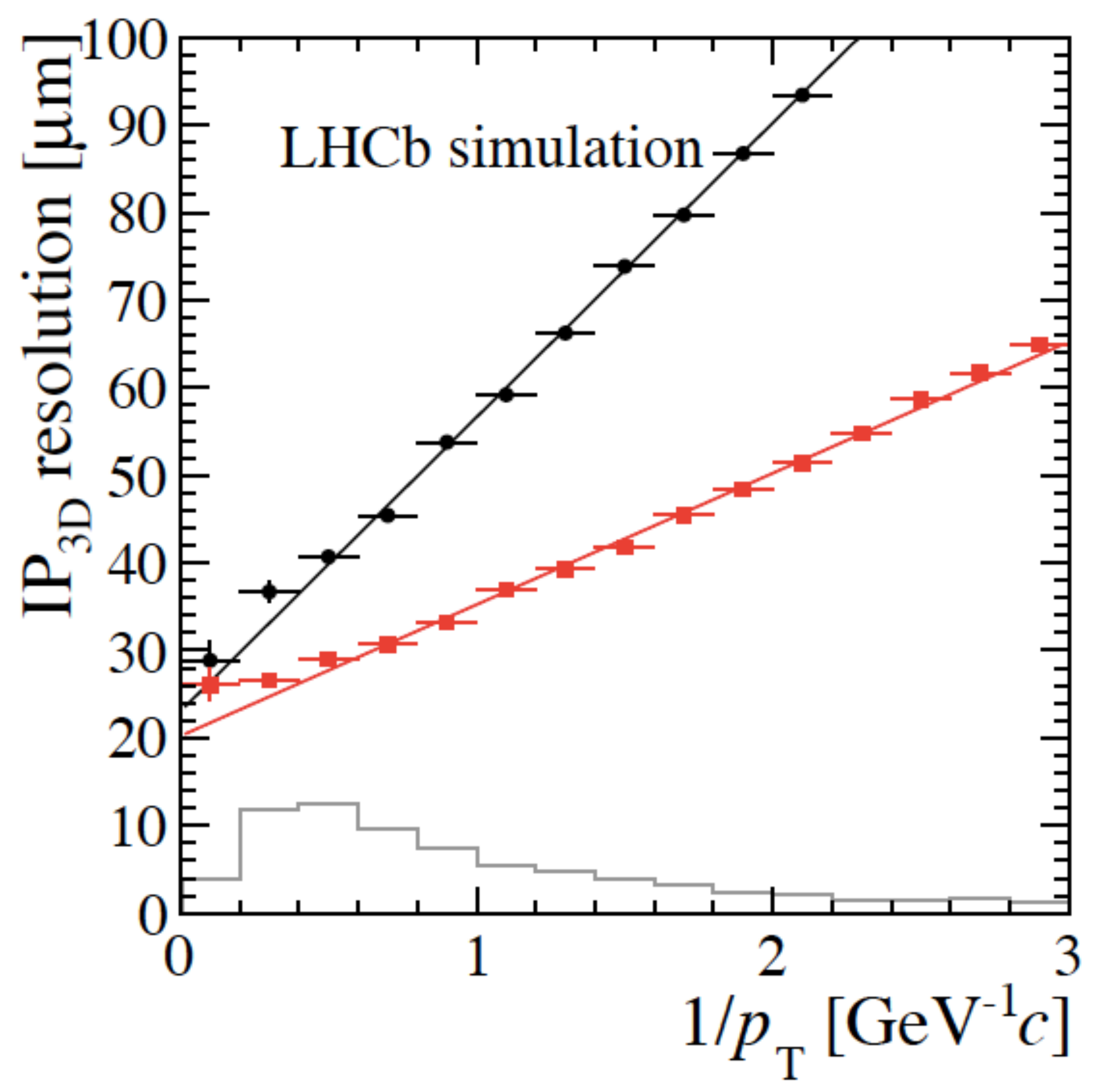}
\includegraphics[width=0.56\linewidth]{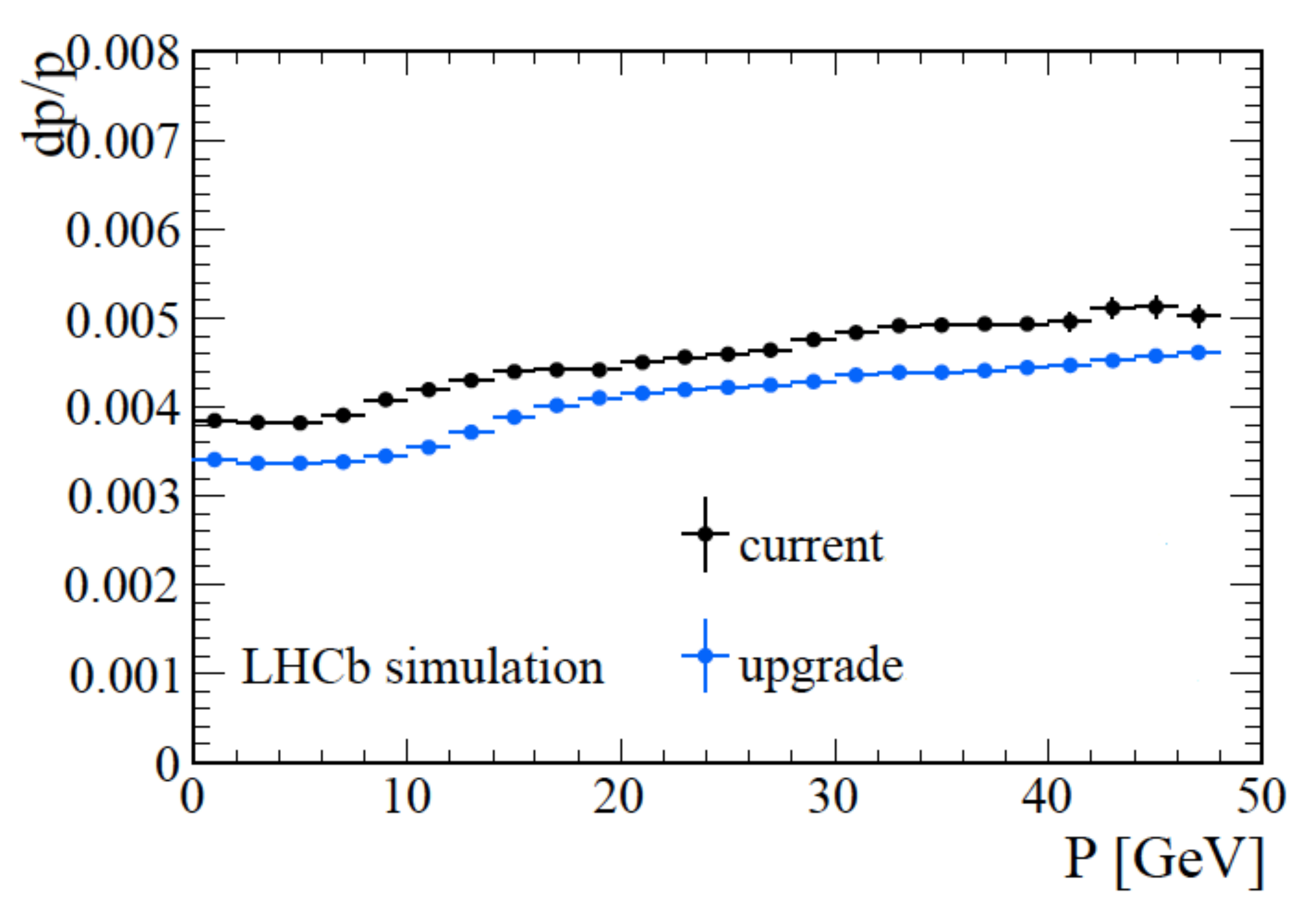}
\caption{ (Left) 3D resolution for the impact parameter (IP) as a function of $\pT^{-1}$, for upgraded (red squares) and current (black squares) VELO layouts. The light grey histogram shows the relative population of b-hadron daughter tracks. 
(Right) Momentum resolution as a function of $p$ for upgraded (blue dots) and current (black dots) tracking layouts.}
\label{FIG1}
\end{figure}

\subsubsection{Particle identification}
Particle identification (PID) is a critical aspect of the LHCb upgrade. Most of the physics reach is determined by the identification of hadrons, muons, electrons and photons. The effort has been focused on maintaining or ameliorating the performance at high luminosity. The detectors involved in this challenge are the RICH, the  CALO and the MUON systems.

The RICH  is essential for the study of hadronic final states, precision measurements of CP violation and rare decays of $b$ and $c$ hadrons. Multibody final states 
such as $B^0_s \to \phi \phi$, $B^0$ and $B^0_s$ decays to $K^+ K^-$ and $\pi^+ \pi^-$ 
would have severe background without $K$ separation.
Moreover, PID is crucial in $b$ tagging and also in the identification of protons from decays of heavy baryonic states.

The changes of geometry (RICH1) and of photon sensor (RICH1 and 2) provide better yield and angular resolution.
The occupancy determines the PID quality, and the most illuminated region of RICH1 shows a manageable 24\%, which is 30\% smaller than achieved with the old geometry at high luminosity.

The PID performance is defined in terms of the efficiency of identifying a true $K$ as a $K$ and the 
mis-identification probability of a true $\pi$ to be tagged as a $K$ or heavier particle.  Simulations give a 2~\% $\pi$ mis-identification efficiency, when a 90~\% $K$ efficiency is requested, comparable with the current performance.

The higher luminosity will lead to a degradation of energy and position measurements for calorimeter objects and to an increase of mis-identification for muons. To mitigate these effects, new algorithms have been developed. The inner muon detectors will be better protected from particles originating from the beam pipe by increasing the shielding.
In the upgrade, at a selection efficiency of 90~\%, mis-identification values of $\sim$1~\% and of $\sim$3~\% are obtained, respectively, for electrons and muons with $p> 10$~GeV, close to the current performance. 

\subsubsection{DAQ and trigger}
The acquisition and the trigger-less software processing of the 40~MHz rate events from LHC represent the keystone of the entire LHCb upgrade.  
The limitations in CPU budget, determined by the size of the Event Filter Farm (EFF), and the output bandwidth, which is constrained by online computing resources, represent the challenges.

Track reconstruction can be performed at close to offline quality with the full input rate and it can be obtained without intermediate selections. The all-software trigger offers
maximal flexibility in designing selections, lowers the $\pT$ thresholds in hadronic events, increasing the efficiency for multi-body final states and for charm events.

Reconstruction time budget and yields for various benchmark channels are the  parameters to define the performance of a  trigger-less scheme. Considering the available CPU power of the EFF at the start of the upgrade, the maximum processing time per event is estimated to be 13~ms, which is considered to be a safe value for event reconstruction.

The gain in efficiency has been studied for several channels and depends strongly on the allowed bandwidth; the lower the bandwidth, the more stringent must be the cuts to stay within allowed limits. 
A conservative approach is adopted, with an accepted HLT output at 25~kHz.
An example of a channel with boosted efficiency is $B_s \to \phi \phi$ , with 
an increase of a factor 4 in yield.
Considerable gains are obtained for semi-leptonic $B$ decays, up to $\sim$50~\%, depending on the charm decay final state configuration, for three-body charmless $B$ decays and for hadronic open charm $B$ decays (e.g.\ $B^+ \to D^0 [K_s \pi \pi] K^+$), up to a factor of 2 to 3.

\section{Physics Prospects}

The HL-LHC physics program is designed to address fundamental questions about the
nature of matter and forces at the subatomic level. 
Although the Higgs boson has been discovered, its properties can
be evaluated with much greater precision with a ten times larger data set
than the original design goal of 300~\ifb. The low value of the Higgs boson mass
poses the so-called hierarchy problem of the SM, which might be explained
by new physics and from a better understanding of electroweak symmetry breaking.
The imbalance between matter and anti-matter in the universe is the 
big open issue for flavour physics. Finally, there may be a new weakly 
interacting massive particle to explain the existence of Dark Matter.
If evidence of deviations from the SM, including new particles, 
are seen before the Phase-2 upgrade, the HL-LHC will allow further
scrutiny of the new landscape of particle physics. In the 
absence of any such hint, the ten-fold increase in data will
push the sensitivity for new physics into uncharted territory.

\subsection{Higgs Boson Measurements}

Over 100 million Higgs boson
events will be produced in each of the ATLAS and CMS detectors.
The dominant production mode is gluon-gluon fusion (ggF), 
with vector-boson fusion (VBF) a factor of 10 lower,
followed by $WH$ and $ZH$ associated production, and
over a million $\ttbar H$ events. This will 
allow the study of many Higgs boson production and decay modes,
including 20 thousand events in the
golden four-lepton final state. Figure~\ref{fig:HIG} (left) shows the
reconstructed mass distribution of the 4-lepton system for a ggF enriched sample.

The cross section times branching ratio can be 
related to the production cross section and the partial and
total Higgs boson decay widths via the narrow-width approximation:
\begin{equation}
\left(\sigma\cdot{\text{BR}}\right)(\mathit{x}\to H \to\mathit{ff}) = \frac{\sigma_{\mathit{x}}\cdot\Gamma_{\mathit{ff}}}{\Gamma_{\mathrm{tot}}},
\end{equation}
where $\sigma_{\mathit{x}}$ is the production cross section through the
mode $\mathit{x}$, $\Gamma_{\mathit{ff}}$ is the partial decay width into
the final state $\mathit{ff}$, and $\Gamma_{\mathrm{tot}}$ is the total
width of the Higgs boson. 
The coupling to gluons and photons is via a loop, so the quantities
$\sigma_\mathrm{ggF}$, $\Gamma_{\mathrm{gg}}$, and
$\Gamma_{\gamma\gamma}$ are 
sensitive to the possible contributions from new particles.
Measuring associated production of $\ttbar H$ allows a direct
measurement of the top-Higgs Yukawa coupling, rather than 
inferring it from the gluon and photon loops.
Since the Yukawa coupling is proportional to the fermion
mass, the top coupling is expected to be of order one, and
is therefore of particular interest.
The possibility of Higgs boson decays to BSM particles,
with a partial width $\Gamma_{\mathrm{BSM}}$, is
accommodated by expressing $\Gamma_{\mathrm{tot}}$ as 
$\Gamma_{\mathrm{tot}} = \sum \Gamma_{ii} +
\Gamma_{\mathrm{BSM}}$, where the sum is over all SM decays.
The partial widths are proportional to the square of the effective Higgs boson
couplings to the corresponding particles.
To test for possible deviations in the data from the expected SM rates in the different channels,
scale factors $\kappa_{i}$ are introduced to modify each coupling, and constrained from
combined fits to the data~\cite{LHCHiggsCrossSectionWorkingGroup:2012nn}.
Table~\ref{tab:cfit} shows the uncertainties
in $\kappa_{i}$ for integrated datasets
of 300~\ifb\ and 3000~\ifb~\cite{CMS:2013xfa,ATL-PHYS-PUB-2014-016}. The ranges reflect
varying assumptions on the evolution of experimental systematic and theoretical
uncertainties. Progress is needed also in reducing theoretical uncertainties to fully benefit from the potential of high luminosity as demonstrated by these projection studies. These concern both missing higher order calculations, and improvement of parton distribution function knowledge. 
  The expected precision improves from $5$--$15\%$ to $2$--$10\%$ with the HL-LHC.

\begin{table}[h]
\caption {The range of precision (68\%CL) under different assumptions
  on the evolution of systematic uncertainties of the fitted values of $\kappa_\gamma$, $\kappa_W$, $\kappa_Z$, $\kappa_g$, $\kappa_b$,
 $\kappa_t$ and $\kappa_\tau$. }
\centering
\begin{tabular} {l|c|c|c|c|c|c|c}
L (\ifb)  &   $\kappa_\gamma$ & $\kappa_W$ & $\kappa_Z$ &
$\kappa_g$ & $\kappa_b$ & $\kappa_t$ &$ \kappa_\tau$ \\ \hline
300                ATLAS & [9, 9] &  [9, 9] &  [8, 8] &  [11, 14] &  [22, 23] &
[20, 22] &  [13, 14]  \\ \hline
300                CMS & [5, 7] &  [4, 6] &  [4, 6] &  [6, 8] &  [10, 13] &
[14, 15] &  [6, 8]   \\ \hline
3000              ATLAS & [4, 5] &  [4, 5] &  [4, 4] &  [5, 9] &
[10, 12]   &  [8, 11] &  [9, 10]  \\ \hline
3000              CMS & [2, 5] &  [2, 5] &  [2, 4] &  [3, 5] &
[4, 7]   &  [7, 10] &  [2, 5]  \\
\end{tabular}
\label {tab:cfit}
\end{table}

The HL-LHC will give unique access to rare Higgs boson decays. The $Z \gamma$ final
state tests the loop structure of the theory, with the possibility of $4 \sigma$ significance
despite the challenging background. The coupling to fermions is proportional to mass,
so the rate of decays to second-generation fermions is an important test.
With an expected 
branching fraction of $2.2 \times
10^{-4}$, the decay $H \rightarrow \mu \mu$ can be unambiguously observed,
and the projected precision of the Higgs boson coupling to muons is about 5\%.

The Higgs boson might also couple to dark matter candidate particles.
The branching ratio to invisible decays can be tested with the coupling fits.
The resulting expected 95\%~CL limit on this branching fraction is about 10\%. This 
complements direct searches for invisible decays with similar sensitivity.

Studies of Higgs boson pair production at the HL-LHC could provide
insight into  Higgs boson trilinear coupling~\cite{Baglio:2012np}. 
This measurement directly probe the 
Higgs field potential since the self-coupling is related to the derivative of the Higgs potential at its minimum, and is sensitive to new physics processes.
The dominant Higgs boson pair production mode at LHC is through gluon fusion,
with destructive interference between the box diagram contribution
and the process involving the Higgs boson self-coupling.
The cross section is near the minimum for the SM value of the self-coupling,
and increases by about a factor two if the self-coupling is zero.
Initial projections have been made with the final states
$bb\gamma\gamma$, $bb\tau\tau$, and $bbWW$, where the $W$ bosons
decay leptonically. Current ATLAS and CMS studies 
expect about a $2 \sigma$ significance with the full HL-LHC
dataset, resulting in a cross-section uncertainty of about
50\%~\cite{ATL-PHYS-PUB-2014-019, ATL-PHYS-PUB-2015-046, CMS-PAS-FTR-15-002}. Figure~\ref{fig:HIG} (right) shows the projection on the diphoton mass
of a two dimensional fit to the $bb$ and $\gamma\gamma$ masses for one representative
toy experiment in the $bb\gamma\gamma$ channel, which is the
channel giving the best sensitivity.
Further improvements are being pursued and additional final
states are under study.

\begin{figure}
\includegraphics[width=0.42\linewidth]{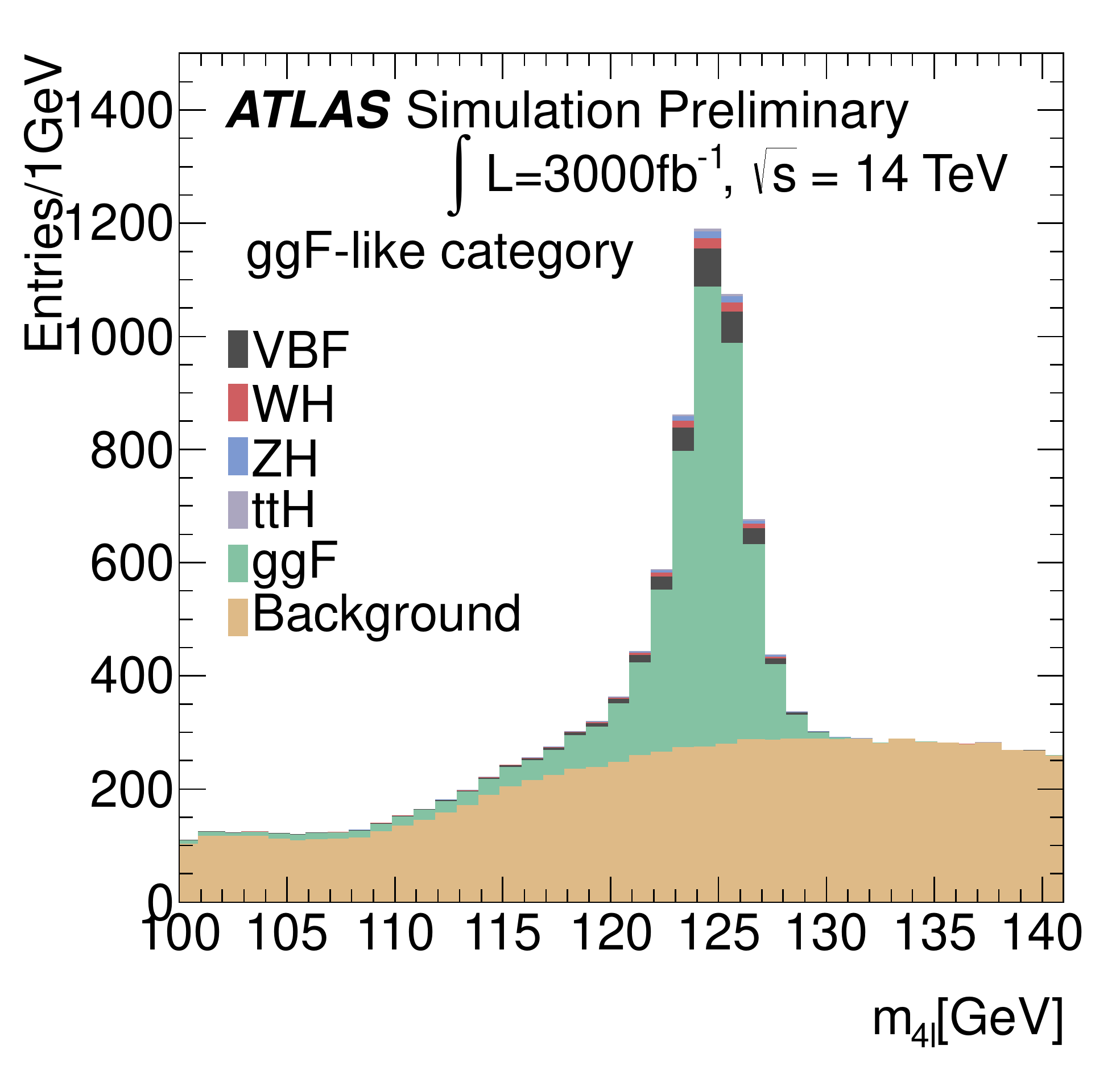}
\includegraphics[width=0.56\linewidth]{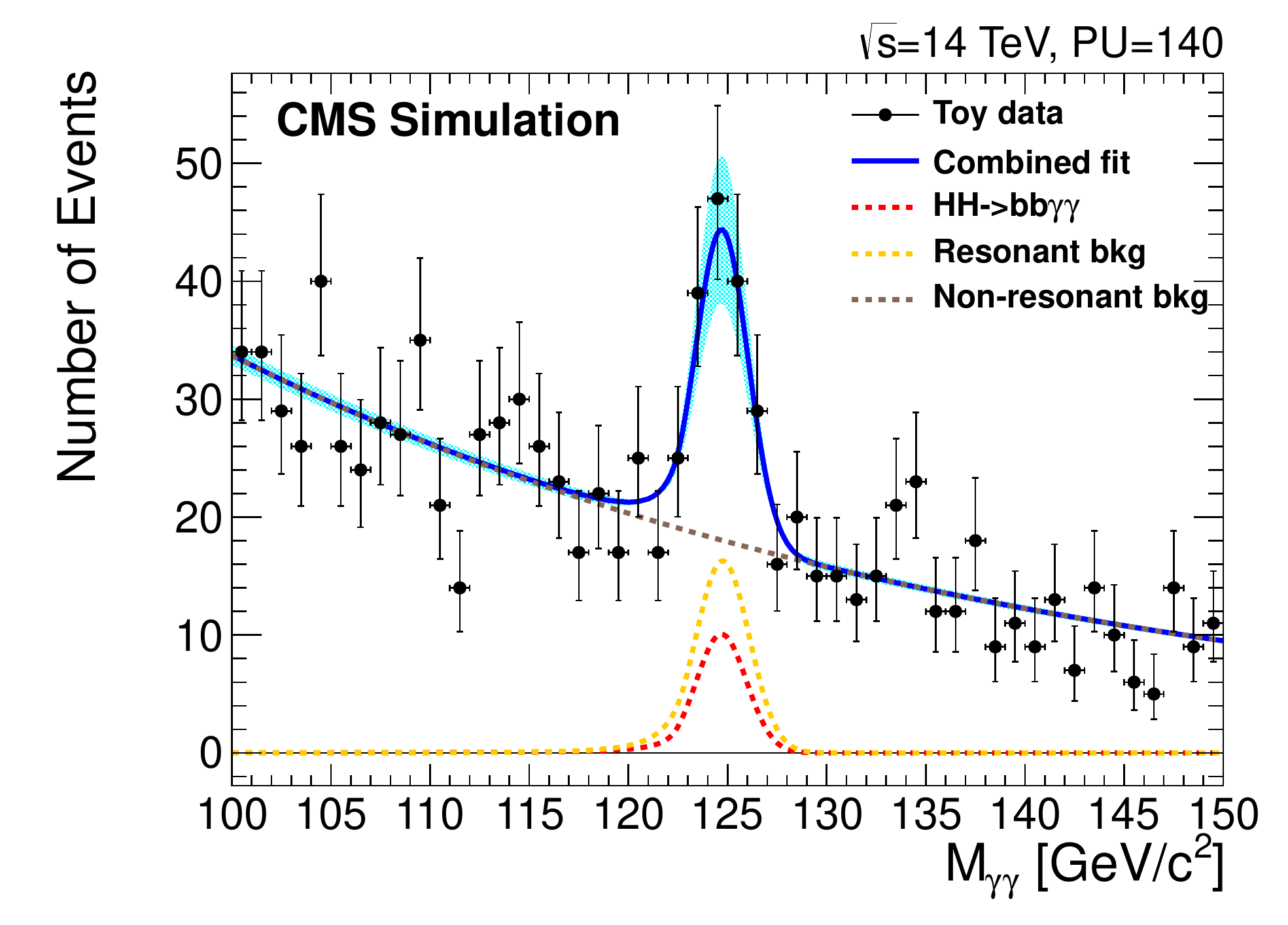}
\caption{ (Left) Reconstructed mass distribution of the 4-lepton system for the ggF-like categories. (Right) Diphoton mass distribution for the
  $bb\gamma\gamma$ channel. The Higgs pair production signal, the resonant background single Higgs boson, and the non-resonant background components are shown in  red, yellow, and gray, respectively.}
\label{fig:HIG}
\end{figure}

\subsection{Standard Model Tests}

Many other investigations of the SM will continue in parallel 
with Higgs boson measurements. For example, vast numbers of $\ttbar$ and single-top events
will allow detailed differential measurements and help to constrain 
PDFs for use in other measurements. Searches for flavour changing neutral
current top quark decays will approach the sensitivity of some BSM models.

Vector boson scattering (VBS) and quartic boson couplings are features
of the SM that remain largely unexplored by the LHC
experiments.  The observation of a Higgs
boson, in accordance with a key prediction of the SM, motivates further study of
the mechanism of electroweak symmetry breaking through measurements of
VBS processes, which share many experimental features with Higgs boson VBF production.
In the absence of the SM Higgs boson, the amplitudes
for these processes would increase as a function of centre-of-mass
energy and ultimately violate
unitarity~\cite{PhysRevLett.38.883,PhysRevD.16.1519}.  The Higgs boson
actually observed by the LHC experiments may restore the unitarity,
although some scenarios of physics beyond the SM predict enhancements
for VBS through modifications to the Higgs sector or the presence of
additional resonances.

The production of same sign $WW$ or $WW$ vector boson pairs can occur through a variety of mechanisms including
double triple gauge coupling (TGC) interactions in the t- or s-channel,
quartic gauge coupling (QGC) interactions, t-channel Higgs boson
exchange and s-channel Higgs boson production. Any addition to the
scattering process would alter the cancellation from the strong
interference of the SM processes, resulting in changes to the cross section at high scattering centre of mass energy.  

Studies of the fully leptonic final states of the same sign $W$ bosons
and $WZ$ boson pairs show that the precision in the measurement of the
electroweak component of the cross section can be determined at the
order of 10\% or better.

\subsection{Beyond the Standard Model}

In most extensions of the SM, including
SUSY and theories with compact extra dimensions,
new particles are expected at the TeV scale.
To date, no such particles have been observed. 
This implies that their  mass is  above the current level of sensitivity, 
that their production rates are lower than expected, 
or that their experimental signatures are very difficult to observe. 
Strongly produced states with large cross sections benefit
most quickly from an increase in centre-of-mass energy,
while the sensitivity for lower cross-section final states, including those
produced by electroweak interactions, grows more
steadily with increasing integrated luminosity.

Supersymmetry is one of the best motivated extensions of the SM, 
controlling the Higgs boson mass by cancelling contributions 
from SM particles and their heavier SUSY partners, a 
natural description of the unification of forces, and
a plausible dark matter candidate.
The huge HL-LHC dataset will allow the investigation of
electroweak production of SUSY particles, for example the
pair production of a chargino and neutralino, $\charg \ninotwo $, decaying to
$ W \neut Z \neut$ or $W \neut H \neut$, yielding characteristic final 
states with three leptons, or a lepton and $b$-jets, together with moderate 
\MET~\cite{ATL-PHYS-PUB-2014-010,ATL-PHYS-PUB-2015-032, CMS-PAS-SUS-14-012}.
The HL-LHC promises a 50 to 100\% increase in mass reach, as shown 
in Table~\ref{tab:susy} for simplified models.

\begin{table}
\centering
\caption{Discovery range ($5\sigma$) for selected simplified SUSY models, where the searches probe
up to the quoted mass.}
\begin{tabular}{l|c|c}
L [\ifb] & 300  & 3000  \\
\hline
$\charg$ [GeV] in $ \charg \ninotwo \rightarrow W \neut Z \neut$ ATLAS & 560 & 820 \\
$\charg$ [GeV] in $ \charg \ninotwo \rightarrow W \neut H \neut$ ATLAS & $< 5 \sigma$ & 800 \\
$\charg$ [GeV] in $ \charg \ninotwo \rightarrow W \neut Z \neut$ CMS &  600 & 900 \\
$\charg$ [GeV] in $ \charg \ninotwo \rightarrow W \neut H \neut$ CMS &  460 & 950 \\ \hline         
stop [TeV] in stop pair production ATLAS & 1.0 & 1.2 \\
sbottom [TeV] in sbottom pair production ATLAS & 1.1 & 1.3 \\
gluino [TeV] in gluino pair production, decay to stop CMS & 1.8 & 2.2 \\ \hline
light squark [TeV] pair production ATLAS & 2.6 & 3.1 \\
gluino [TeV] pair production, decay to quarks ATLAS & 2.0 & 2.4 \\
gluino [TeV] pair production, decay to quarks CMS & 1.9 & 2.2 \\ 
\end{tabular}
\label{tab:susy}
\end{table}

Naturalness motivates SUSY mass spectra with light third generation scalar quarks, stop and sbottom,
to cancel the large top (and bottom) quark contributions to the Higgs boson mass. 
The stop and sbottom may be pair produced directly, or arise in a gluino decay 
chain~\cite{ATL-PHYS-PUB-2013-011,  CMS-PAS-FTR-13-014}.
The final states have multiple top or $b$-quark decays, leading to multiple jets, $b$-tagged jets, possibly one or more leptons
and \MET.
An example of the exclusion and discovery potential for such a search
is shown in Figure~\ref{fig:SUSY} (left)~\cite{ATL-PHYS-PUB-2013-011}.
The reach of searches for strongly produced pairs of 
first and second generation squarks or gluinos will also be extended. Final 
states range from a single high-$\pT$ jet recoiling against missing
transverse energy to signatures with very high jet multiplicity.
Table~\ref{tab:susy} also includes the results of several simplified models
for strongly produced SUSY particles, with typical gains of 20\% in mass reach.

While each search targets specific processes and the
experimental signature, in many cases the signatures are relevant to
multiple SUSY processes. This has been demonstrated in a combined
study of five full-spectrum SUSY models~\cite{CMS-PAS-SUS-14-012},
applying the results of nine distinct SUSY searches, as illustrated
in Figure~\ref{fig:SUSY} (right).  The first three models are motivated by naturalness~\cite{Papucci:2011wy}
and differ by the mass of the sleptons and also by the composition of neutralinos and charginos, which are mixtures of binos, winos, and higgsinos.
Two other models contain coannihilation scenarios, which are motivated by their prediction 
of the relic density.

\begin{figure}
\includegraphics[width=0.49\linewidth]{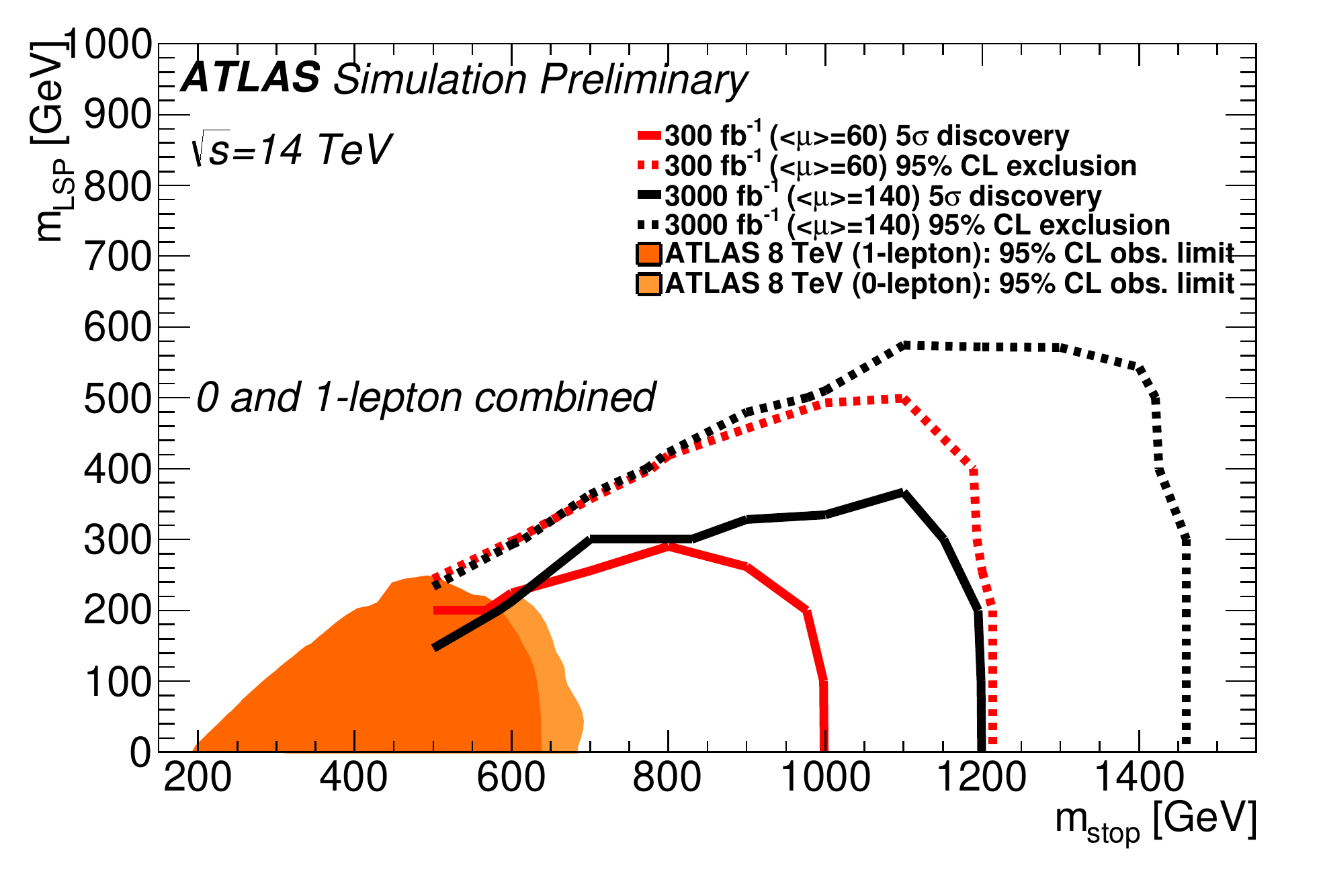}
\includegraphics[width=0.49\linewidth]{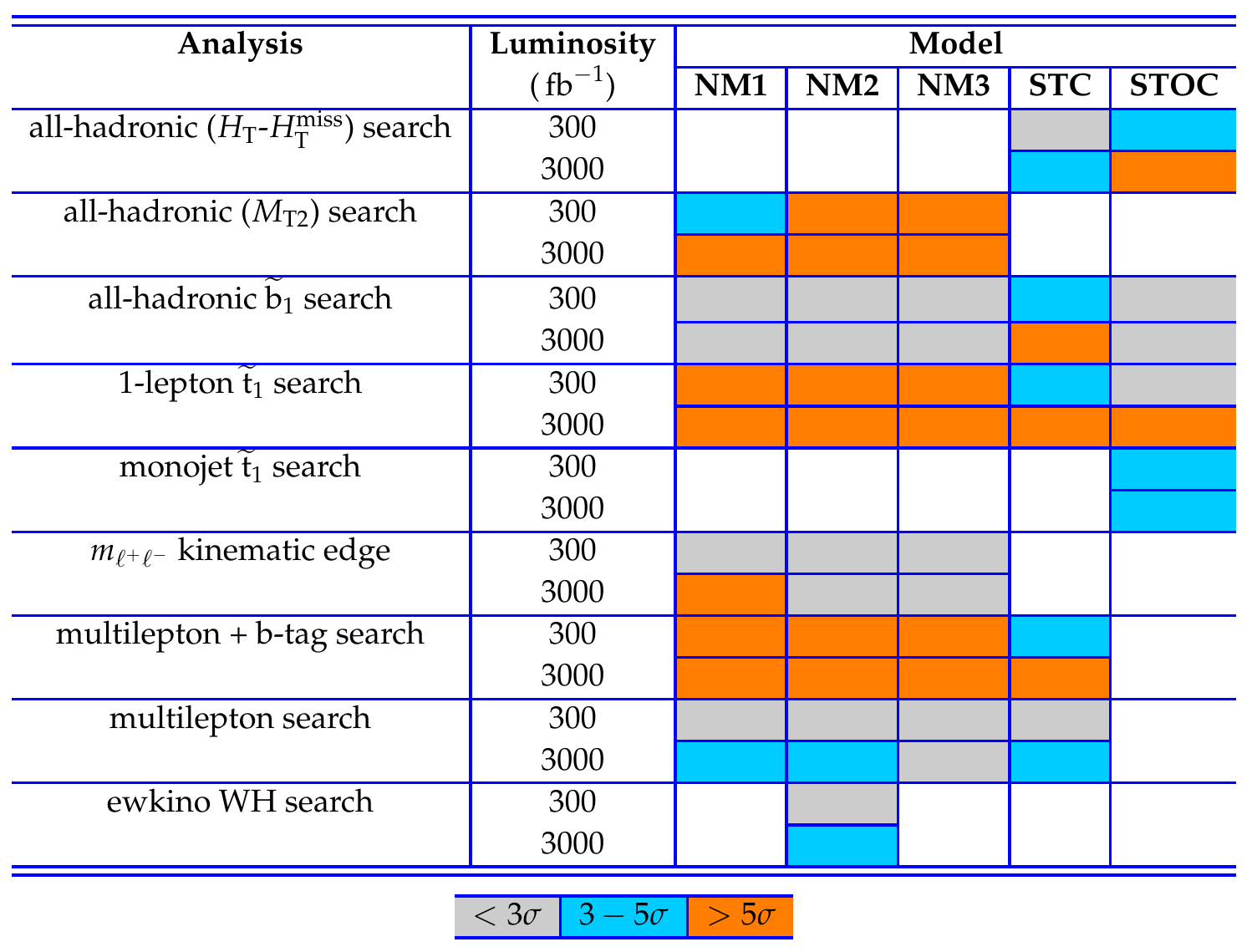}
\caption{ (Left) The 95 \% CL exclusion limits (dashed) and 5-sigma
discovery reach (solid) for 300\ifb (red) and 3000 \ifb (black) in the
stop-$\neut$ mass plane assuming $\tilde{t} \rightarrow t \neut$.
(Right) Overview of nine distinct SUSY analyses and their application
to five full-spectrum benchmark models. Varying patterns of
discoveries arise in different final states.}
\label{fig:SUSY}
\end{figure}

Beyond SUSY, a large number of models predict a variety of signatures accessible at
the LHC. Resonances are straightforward targets for searches for
new physics. Narrow resonances such as the neutral $Z '$ with varying
branching ratios in different models
are predicted by Grand Unified Theories (GUT)~\cite{Langacker:2008yv}.
The Randall-Sundrum (RS) model~\cite{Randall:1999ee} of small
extra dimensions also predicts additional resonances, such as Kaluza-Klein gluons.
Dark matter is one of the
most promising of many hints for physics beyond the SM. The LHC probes
various types of interactions and dark matter can be detected by
tagging SM particles in the detector and identifying the recoiling dark matter particles
from a missing transverse energy signature. Heavy stable charged
particles (HSCP) create even more exotic signatures in the detectors. HSCP candidates are
non-relativistic and yield anomalous energy loss measurements or large
time-of-flight. A summary of the discovery reach and exclusion limits for selected non-SUSY
models of new physics is shown in
Table~\ref{tab:exo}~\cite{ATL-PHYS-PUB-2013-003, CERN-LHCC-2015-010}. The ten times
larger dataset of the HL-LHC yields approximately a 20\% increase in mass reach.
Furthermore, the HL-LHC dataset will allow detailed studies of the
newly discovered particles, if hints of new physics signals are seen by the end of Run 3.

\begin{table}
\centering
\caption{Discovery range and 95\%~CL exclusion limits for selected
  non-SUSY models of new physics, where the searches probe up to the quoted mass in TeV.}
\begin{tabular}{l|c|c}
$5\sigma$ discovery L [\ifb] & 300  & 3000  \\
\hline
Z' SSM [TeV] in ee + $\mu\mu$ CMS & 5.2 & 6.4 \\
Z' Psi [TeV] in ee + $\mu\mu$ CMS & 4.6 & 5.7 \\
W' SSM [TeV] in e$\nu$ CMS & 4.8 & 5.9 \\
Dark Matter in e$\nu$ xi=-1 ($\Lambda$ [TeV]) CMS & 2.6 & 3.7 \\
Dark Matter in e$\nu$ xi=0 ($\Lambda$ [TeV]) CMS & 1.8 & 2.4 \\
Dark Matter M* ATLAS  & 2.2 & 2.6 \\
HSCP gluino [TeV], coupling f=10 CMS & 2.1 & 2.5 \\
HSCP direct pair-produced stau [TeV] CMS & 0.6 & 1.1 \\
T-quark [TeV] pair production CMS & 1.3 & 1.6 \\ \hline
95\% CL exclusion limit & & \\ \hline
Z' SSM [TeV] in ee ATLAS & 6.5 & 7.8 \\
Kaluza-Klein gluon [TeV] in $\ttbar$ ATLAS & 4.3 & 6.7 \\
\end{tabular}
\label{tab:exo}
\end{table}

\subsection{Flavour physics}

B factories and LHC data demonstrated the success of the Cabibbo-Kobayashi-Maskawa (CKM) paradigm. All the measurements agree in a highly precise and profound way and
the quark flavour sector is well described by the  CKM mechanism. 
The gain in statistical precision from high luminosity operation will impact a large set of variables, measured not only by the dedicated flavour experiment, LHCb, but also by the general purpose ATLAS and CMS detectors, profiting from their large integrated luminosity. 
LHCb data from Run~2 ($\sim 5$~\ifb), will increase the sample by a factor 4 with respect to the present one, and should confirm or discard the anomalies mentioned above.
In the lucky case that any of these effects is still there at the end of Run~2,  the upgrade would open interesting scenarios.
Alternatively, an increase in statistics by a factor ten or more  will probe some of the flavour variables at a precision not far from their theoretical uncertainties. 

The perspectives for the very rare leptonic decays  ($B_{d,s} \to \mu \mu$),
the mixing induced CP violation in $B^0_s$ decays and the tree-level determination of the CKM angle $\gamma$ are presented here. 
Extrapolations for LHCb assume that the upgraded detector will operate at high luminosity with a performance similar to the present configuration, while they do not include, conservatively, any increase in statistics due to the use of lower $\pT$ thresholds.

\subsubsection{Very rare $B_{d,s}$ decays}
The decays $B_{d,s} \to \mu \mu$ are a special case amongst the electroweak penguin processes, as
they are chirality-suppressed in the SM and are most sensitive to scalar and pseudoscalar operators.
Therefore, in the SM these decays are exceedingly rare, with predicted decay rates for $B^0_s$ and $B^0$ of $3.6  \times 10^{-9} $ and $1.1 \times 10^{-10}$, respectively, and $\sim$ 10\% theoretical uncertainty~\cite{bobeth}.
New physics can produce significant enhancements in the decay rates.
The ratio of branching fractions for $B^0_s$ and $B^0$ is known with better theoretical uncertainty and  also provides a stringent test of Minimal Flavour Violation (MFV) models.
In Run 1 CMS and LHCb have observed  the decay rate of the $B^0_s \to \mu \mu$ channel finding it in good agreement with the SM, and significantly restricting the available phase space for BSM theories, in particular for MSSM models. 
With increased statistics the determination of the effective
lifetime and time-dependent CP violation could also become possible \cite{debruyn}.
The effective $\tau_{\mu \mu}$ may still reveal new physics effects, even if 
the $B^0_s \to \mu \mu$  decay rate agrees with SM expectations.
 
Prospects for more precise measurements at HL-LHC are interesting, although
they will still be dominated by the experimental uncertainty. 
In CMS, these  channels are benchmarks~\cite{CMSb} for assessing the B-physics performance of the HL-LHC layout. Two upgrades are relevant for the analysis, the implementation of a Level-1 track trigger and the increased momentum resolution thanks to reduced material budget in the tracker. 
A low $\pT$ di-muon Level-1 trigger can reduce the rate of interesting events to a few hundred Hz. The resolution in invariant mass   in the barrel decreases to $\sim 30$~MeV, 50\% better than the current one, see Figure~\ref{FIG2} (left). 
The sensitivity for the branching fractions of $B^0$ and $B^0_{s}$ can be measured in CMS with a
precision of 18\% and 11\% , respectively. 
Similar studies have been also performed for LHCb~\cite{implications}, showing that the determination of the ratio of branching ratios has 35\% accuracy with $\sim 50$~\ifb.

\subsubsection{$B^0_{s}$ mixing measurements}
The CP-violating phase $\phi_s$ arises in the interference between the amplitudes of $B^0_s$
mesons decaying via the $b \to c \overline{c} s$ tree diagram to CP eigenstates directly and those
decaying after oscillation. 
In the SM, a global fit to experimental data leads to an expected value of  
$\phi_s$ = (-0.0363 $\pm$ 0.0013) rad, see Figure~\ref{FIG2} (right).
Non-SM particles could contribute to $B^0_s$
oscillations and a measurement of $\phi_s$  different from the prediction
would provide unambiguous evidence for new physics.
The $B^0_s \to J/\psi K K$ and $B^0_s \to J/\psi \pi \pi$ decays are the channels providing the best sensitivity and at high luminosity the final experimental precision should allow the observation of
changes as small as a factor of two with respect to the SM  with 3 $\sigma$ significance. 

The current LHCb experimental uncertainty on $\phi_s$ is 0.050 rad and an extrapolation to $50$~\ifb\ shows that a value of 0.008 rad can be reached.
It is expected that the main limitation will be statistical, and the present largest systematic uncertainties  (background description, angular acceptance) are expected to decrease with more sophisticated analyses or to scale with statistics. 
A case study  has been done also for the ATLAS experiment~\cite{ATLASPhis}, where the potential will be strongly driven by trigger thresholds. 
The upgraded tracking system will improve the decay time resolution by 30\% , and
a total uncertainty as low as 0.02 rad is expected with the full HL-LHC luminosity. Possible improvements could come from the Level-1 topological trigger and the  use of flavour tagging.

Moreover, the study of the penguin-only mediated $b \to s q \overline{q}, (q=s;d)$ decays will probe
the presence of physics beyond the SM that can be detected by looking for its contribution to
channels mediated only by loop diagrams. 
Among these decays, $B^0_s \to \phi \phi$ provides a rich set of observables that are
rather precisely known in the SM but could be affected by
new heavy particles appearing in the loop.
The variable $\phi_s$ measured here will be compared with the tree level determinations. 
LHCb can reach an accuracy comparable to the theoretical indetermination (0.02 rad) by the end of high luminosity run.

\begin{figure}
\includegraphics[width=0.44\linewidth]{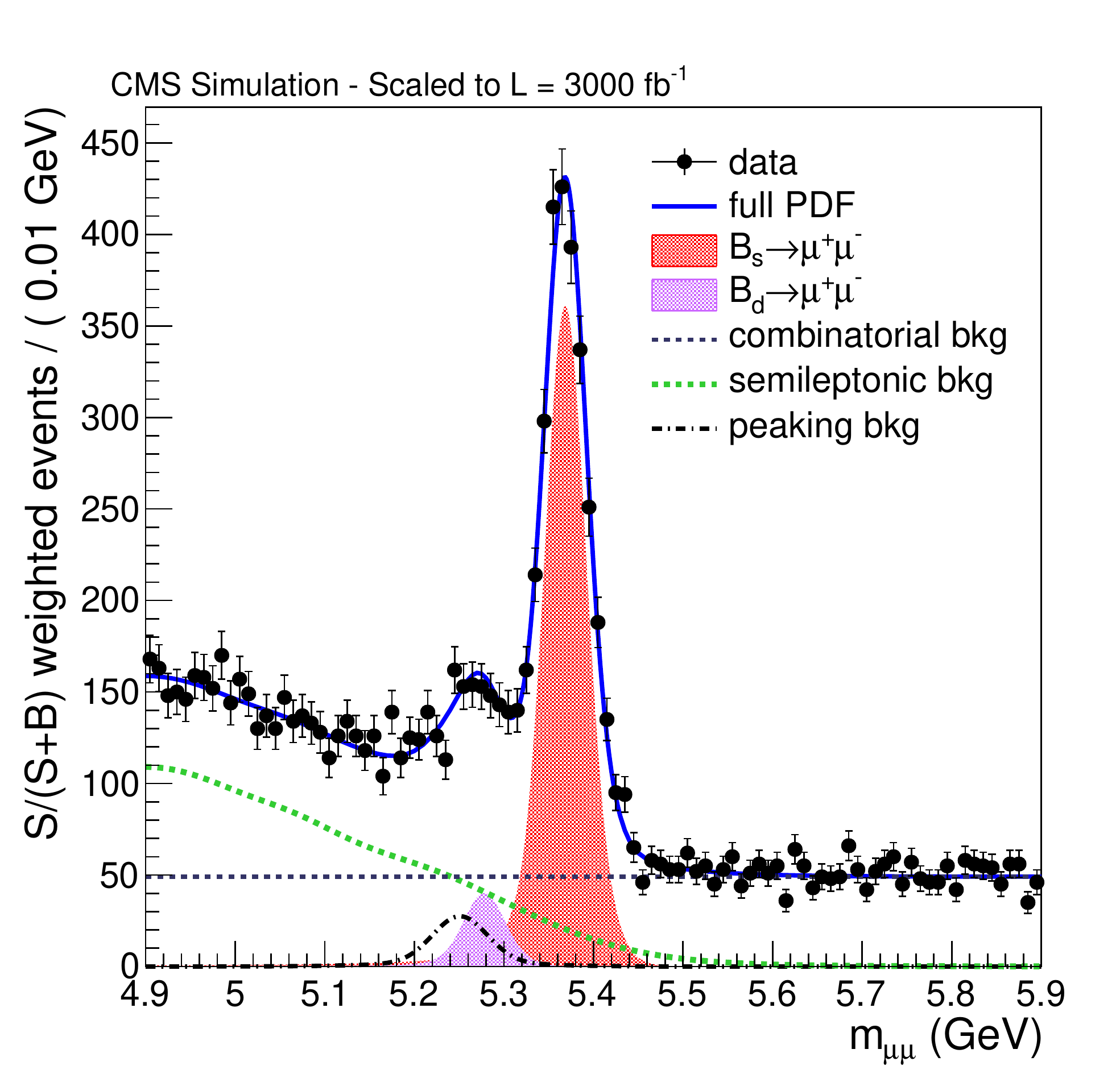}
\includegraphics[width=0.54\linewidth]{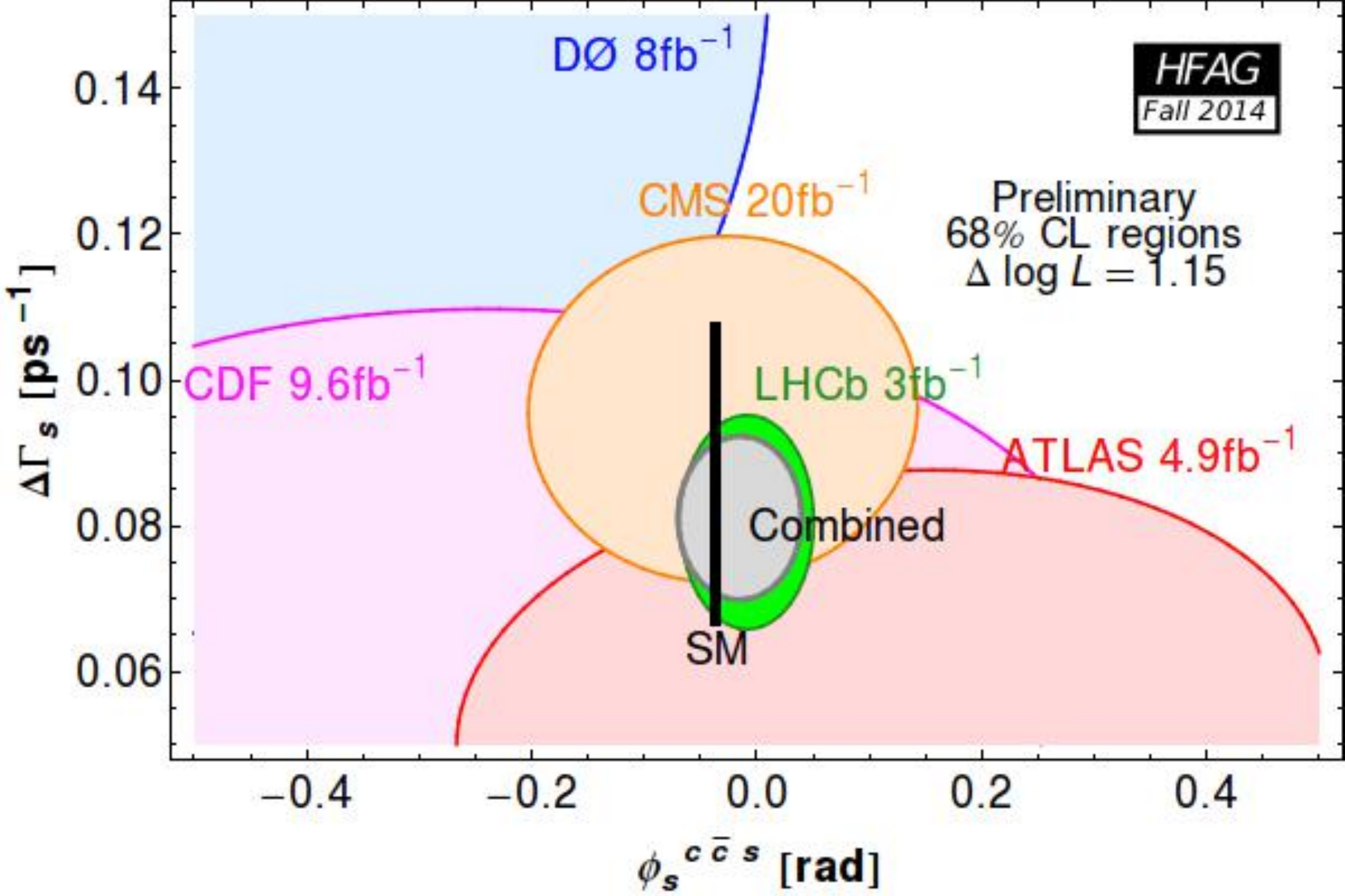}
\caption{ (Left) Projections of the $B_{d,s} \to \mu \mu$ mass fits to 3000~\ifb\ of integrated
luminosity for the upgraded CMS detector (barrel only). (Right) The current experimental situation for the determination of $\phi_s$ and $\Delta \Gamma_s$ The SM expectation values are also shown.}
\label{FIG2}
\end{figure}

\subsubsection{Measurement of CKM $\gamma$ angle}
Another important example for the improvement in precision
is the determination of the angle $\gamma$, which is one of the least well known parameters of the CKM matrix. 
It has negligible theoretical uncertainties ($\sim 10^{-6}$) and serves as the reference point for comparison with $\gamma$ value obtained from loop decays (like $\Delta m_d$ and $sin~2\beta$) and to improve the precision of global CKM fits and the corresponding limits on new physics contributions. 
Combining several independent decay modes is the key to achieving the ultimate precision,  with time-independent channels ($B \to DK$) or with time-dependent ones ($B_s \to D_s K$), involving only tree amplitudes. 
Further information can come from charmless decays, although penguin pollution makes theoretical control more difficult.
Given its particle identification capabilities, LHCb is the unique experiment at the LHC which can measure $\gamma$, and at high-luminosity
a global fit over the various channels could provide an estimated precision below $1^{\circ}$. 
BELLE II at KEKB  with a data sample of 50~ab$^{-1}$ could provide  a comparable accuracy.  

\subsubsection{Other measurements: quarkonia and exotic states}
LHC experiments also have excellent potential in a range of other  topics connected to production of particles with heavy quarks.
Quarkonia production studies have been focussed on the
$J/\psi , \psi(2S)$ and $\Upsilon$ decaying to dimuons. 
With increasing luminosity,  access to reactions with low cross sections will become possible, and studies with new probes will be pursued. 
Multiple quarkonia production (such as double $J/\psi$,  $\psi(2S)$ or $\Upsilon$, or combinations) originated by single or double parton scattering, or from central exclusive processes and samples of charmonium and  bottomonium states via hadronic decay modes will become statistically significant.
This will allow more detailed studies of the $X(3872)$ and $Z(4430)$ particles and of exotic bottomonium. 
A large sample of tetra-quark and penta-quark candidates, deriving not only from $B$ mesons but also from $b$-baryon decays ($\Lambda_b, \Sigma_b, \Xi_b, \Omega_b$)  could become available.

\section{Conclusion}

Thanks to the excellent performance of the LHC and of the four experiments (ALICE, ATLAS, CMS and LHCb), the first three years of data taking of the LHC Run 1 (2010-2013) were tremendously successful,
crowned by the headline discovery of a Higgs boson consistent with the expectation of the SM Higgs boson.
Moreover, an impressive array of measurements of SM phenomena have been accomplished, including flavour physics. Apart from a few deviations needing further experimental and theoretical scrutiny, the full picture leads to a scenario in which, despite its theoretical shortcomings, the SM has not shown any defect yet. 

Exploiting the increase in energy and in statistics, LHC Run~2 (2015-2018) will further explore possible deviations from the SM and some preliminary results with limited statistics from 2015 have provoked
intense interest.
An exciting program of measurements is planned with Run~3 (2021-2023), which concludes the original target integrated luminosity of the LHC. Run~3 will also see the start of the data taking of an upgraded LHCb detector operating at an instantaneous luminosity a factor 5 higher than in previous runs.
In the meantime, a huge investment will be made in the HL-LHC project to bring the 
instantaneous luminosity to 5 to 7.5 $\times 10^{34}$~cm$^{-2}$s$^{-1}$.
The start of HL-LHC operation is planned for 2027.

The LHC detectors are undergoing an intense program of upgrades, to be completed in 2026 by the general purpose detectors and in 2020 by ALICE and LHCb, respectively.
The main elements involved in the upgrades are better tracking performance, to cope with the high pile-up conditions, enhanced trigger capabilities and new DAQ and computing infrastructures. 
The detectors will be upgraded to sustain the high luminosity operation and to integrate pp collisions up to 3000~\ifb\ (ATLAS and CMS) in approximately ten years of operation (50~\ifb\ for LHCb). 

The overarching ambition of the LHC experiments is to detect new phenomena that can explain the current theoretical problems of the SM. This might come in the form of new particles, or deviations from the SM prediction in precision measurements of Higgs boson properties, or other electroweak, QCD or flavour observables. The HL-LHC data will enable the improvement of the precision on Higgs boson couplings by a factor of 2 to 3 with respect to the LHC. The mass reach to discover new particles increases by about 20\% for strongly produced processes and in some case by 50-100\% in weak production modes.

This challenging program spans more than two decades from now and will be able to exploit fully the many opportunities that the LHC machine offers to particle physicists, hoping to provide answers to the most fundamental questions posed by nature.

\section*{ACKNOWLEDGMENTS}
The authors wholeheartedly thank the LHC team for the outstanding performance of the
machine, and their colleagues in the ATLAS, CMS and LHCb
Collaborations for the productive atmosphere that led to many of the
results and prepared the upgrade plans discussed here. They also
acknowledge with gratitude A.~DeRoeck, V.~Gligorov, and L.~Nisati for careful reading and useful comments on the manuscript.

\bibliography{ar-hllhc.bib}
\bibliographystyle{ar-style5.bst}

\end{document}